\documentclass[11pt]{article}

\usepackage{epsfig, amsfonts, amsmath, amssymb, amsthm, pictex, verbatim, amsthm, pdfsync, mathrsfs}
\usepackage{tensind}
\usepackage{endnotes}
\usepackage[normalem]{ulem}

\usepackage{hyperref}  %This package should often be loaded last, to prevent conflict with other packages.
\hypersetup{colorlinks=true, citecolor=blue, urlcolor=blue,linkcolor=blue} 
\usepackage{graphicx}
\usepackage{amsmath}
\usepackage{cases}
\usepackage{amssymb}
\usepackage{hyphenat}
\usepackage[hypcap]{caption} %When hyperrefs link to figures, they go to the top of the figure, instead of to the caption.

\usepackage{tikz} 					%Drawing pictures within LaTeX.
\usetikzlibrary[patterns]				% Different patterns to fill areas.
\usetikzlibrary[arrows]				% Different arrow heads
\usetikzlibrary{positioning}			% Rotating (e.g. vertical axis labels)
\usepackage[colorinlistoftodos, textwidth=2.8cm, textsize=tiny, shadow, disable]{todonotes}				%Inserting todo notes.  %Change colour of bublles and remove lines: \todo[noline,color=red]{comment}
\usepackage{marginnote}

\usepackage{framed}			%The package creates three environments:framed, which puts an ordinary frame box around the region,shaded, which shades the region, and left­bar, which places a line at the left side. The environments allow a break at their start (the \FrameCommand enables creation of a title that is “attached” to the environment); breaks are also allowed in the course of the framed/shaded matter.

\usepackage{comment}

\usepackage{colortbl} %Allows coloring of single cells of a table via \cellcolor{}
\usepackage{booktabs} % for much better looking tables
\usepackage{multicol}
\usepackage{multirow}
\usepackage{tabularx}
\usepackage{fixltx2e} % Allows for \textsubscript

\usepackage{caption}
\usepackage{subcaption}   % These two packes (caption and subcaption) are used to get nice arrays of figures.

\usepackage{cite}  % [1,2,3]  will become [1-3].

\usepackage[margin=1.52in]{geometry}
\usepackage{authblk}

\usepackage{xparse}

\makeatletter
\newcommand{\customlabel}[2]{ \protected@write \@auxout {}{\string \newlabel {#1}{{#2}{\thepage}{#2}{#1}{}} } \hypertarget{#1}{}}
\makeatother
\definecolor{cblue}{RGB}{100,5,255}    
\definecolor{cred}{RGB}{155,50,40} 
\definecolor{cgreen}{RGB}{5,165,20}  
\definecolor{corange}{rgb}{1.0,0.49,0.0}  
                                                                   
\newcommand{\g}{g_{\mu \nu}}

\title{Cartography of the space of theories: an interpretational chart for fields that are both (dark) matter and spacetime}
\author[1,2,3]{Niels C.M.\ Martens}
\author[1,2]{Dennis Lehmkuhl}
\affil[1]{DFG Research Unit ``The Epistemology of the Large Hadron Collider'' (grant FOR 2063)\newline \url{martens@physik.rwth-aachen.de}, \url{dennis.lehmkuhl@uni-bonn.de}}
\affil[2]{Lichtenberg Group for History and Philosophy of Physics, University of Bonn}
\affil[3]{Institute for Theoretical Particle Physics and Cosmology, RWTH Aachen University\vspace{3mm} \newline \normalfont{This is a postprint (i.e.~post-peer-review but pre-copyedit) version of an article accepted for publication in} \textit{Studies in History and Philosophy of Modern Physics}. \normalfont{The final authenticated version is/will be available online (open access) at: \url{www.sciencedirect.com/science/article/abs/pii/S1355219820301106}.}}
\begin{document}
	
\maketitle

\abstract{This paper pushes back against the Democritean-Newtonian tradition of assuming a strict conceptual dichotomy between spacetime and matter. Our approach proceeds via the more narrow distinction between modified gravity/spacetime (MG) and dark matter (DM). A prequel paper argued that the novel field $\Phi$ postulated by Berezhiani and Khoury's `superfluid dark matter theory' is as much (dark) matter as anything could possibly be, but also---below the critical temperature for superfluidity---as much (of a modification of) spacetime as anything could possibly be. Here we introduce and critically evaluate three groups of interpretations that one should consider for such Janus-faced theories. The consubstantiality interpretation holds that $\Phi$ is both (dark) matter and a modification of spacetime, analogously to the sense in which Jesus (according to catholicism) is both human and god. The fundamendalist interpretations consider for each of these roles whether they are instantiated fundamentally or emergently. The breakdown interpretations focus on the question of whether $\Phi$ signals the breakdown, in some sense to be specified, of the MG-DM dichotomy and perhaps even the broader spacetime--matter distinction. More generally, it is argued that hybrid theories urge a move towards a single space of theories, rather than two separate spaces of spacetime theories and matter theories, respectively.}

\newpage
\tableofcontents

%\vspace{8mm}

%\begin{quote}
%	There simply is not a fact of the matter here as to what should count as ``space'' and what should count as a ``physical object'' in the sense of the original ``space-matter'' distinction that spawned the debate. Present-day physicists do \emph{not} employ a language that conforms with the original contrast
	
%	\hfill --Robert Rynasiewicz \cite[p.301]{rynasiewicz1996}
%\end{quote}

\section{Introduction}

%\todo{mention cartograpy somewhere}

Democritus claimed that everything in our universe is ultimately reducible to atoms (matter) or void (space). This strict conceptual and metaphysical dichotomy between matter and space(time)---everything in our universe fits into exactly one of those categories---has reigned supreme ever since Newton revived it. Recently, several approaches to the problem of quantum gravity have put pressure on these two categories and the strict dichotomy between them. There is however reason to worry long before reaching the region where quantum gravity becomes relevant. (Moreover, focusing on these cases has the additional benefit of staying closer to experiment/observation.) How to categorise the metric in plain old general relativity (GR) is already controversial, as it exhibits both properties that are standardly associated with spacetime as well as properties that are standardly associated with matter \cite{anderson1967,earman1987,maudlin1988,feynman1995,hoefer1996,rynasiewicz1996,rovelli1997,dorato2000,slowik2005,dorato2008,lehmkuhl2008,rovelli2010,lehmkuhl2011,rey2013,read2018,duerr2019a,duerr2019b,duerragainstread,martenslehmkuhl1} \cite[p.354]{vassallo2016} \cite[Ch.9]{brown2005} \cite[\S 3.3]{belot2011b} \cite[\S 8]{greaves2011} \cite[p.36]{lehmkuhl2018}. We contend that the pressure becomes even more acute when considering certain theories featuring in the debate between dark matter (DM) and modified gravity/spacetime (MG). 

In particular, the prequel \cite{martenslehmkuhl1} to this second of two companion papers argued that the novel field $\Phi$ postulated by Berezhiani and Khoury's `superfluid dark matter theory' (SFDM) \cite{berezhiani2016,berezhiani2015} is as much (dark) matter as anything could possibly be, \emph{but also}---below the critical temperature for superfluidity---as much (of a modification of) spacetime as anything could possibly be. This first result leaves open certain interpretational questions. For instance, does the fact that $\Phi$ instantiates the spacetime role only below the critical temperature for superfluidity imply that it is `really' dark matter, acting (in the sub-critical-temperature regime) `as if' it were a modification of gravity and spacetime? Or is the sense in which $\Phi$ is dark matter on a par with the sense in which it is a modification of gravity and spacetime? Does the result of the prequel paper put pressure on the contemporary importance, usefulness or even coherence of the dark matter/ modified gravity dichotomy and perhaps even on the broader matter--spacetime distinction?

It is the aim of this paper to unpack these and other interpretational questions. After introducing `superfluid dark matter theory' in more detail in section \ref{FDMintro} and reiterating the conclusion of the prequel paper in section \ref{spacetime-matter}, we will introduce and critically evaluate nine potential interpretations---distributed across three groups---that we believe one should consider when attempting to understand theories with objects that do not all fall neatly into one of the two categories (i.e.\ spacetime and matter). The goal is not so much to advocate a single interpretation that is definitely to be preferred---we argue that four out of nine interpretations are currently viable for SFDM---but to put on the table a chart of viable ways of understanding theories like SFDM and advantages of and problems with each approach. This is partially because the theory is very new (and, in particular, known to be incomplete, as will be discussed below), and partially because the difference between some interpretations might be best fought out at the level of the space of theories rather than within a single theory. This case study of SFDM serves as a bottom-up approach to the cartography of this space of theories,\footnote{Note that this cartographic project differs from the related cartographic project by Slowik, who aims to provide a map of substantivalist and relationalist positions \cite{slowik2018}.} by inspiring a chart of interpretations that may help us understand and navigate that large space. Section \ref{jesus} introduces the consubstantiality interpretation, which holds that $\Phi$ is both (dark) matter and a modification of spacetime, analogously to the sense in which Jesus (according to catholic doctrine) is both human and god. The (four) fundamendalist interpretations (section \ref{fundamentalist}) consider for each of these roles whether they are instantiated fundamentally or emergently. The (four) breakdown interpretations (section \ref{breakdown}) focus on the question of whether $\Phi$ signals the breakdown, in some sense to be specified, of the MG-DM dichotomy and perhaps even the broader spacetime--matter distinction. Section \ref{cartography} discusses the relations between these interpretations, and reflects upon the lessons learned regarding the DM/MG distinction, the broader spacetime--matter distinction, and the cartography of the space of theories.

\section{`Superfluid Dark Matter theory' (SFDM)} \label{FDMintro}

In order to make this paper somewhat self-contained and in order to be able to refer back to the relevant Lagrangians, we almost verbatim reproduce the introduction to `superfluid dark matter theory' from the previous paper \cite{martenslehmkuhl1}, followed by comments regarding its completeness. Readers familiar with the prequel paper may wish to skip ahead to the final paragraph of this section.

Theories labeled as dark matter theories have traditionally done well at the level of cosmology and galaxy clusters, but less so at the level of galaxies. The opposite is the case for theories labeled as modifications of gravity and/or spacetime. A promising approach to breaking this stalemate is to find a single novel entity for which there is a natural, physical, dynamical reason why it behaves like DM on large scales and like MG on galactic scales. In this respect it is relevant to note that to mediate a long-range force in galaxies, a massless messenger (force carrier) is needed. A natural candidate presents itself: the quantised soundwaves of a superfluid, i.e.\ phonons, which are Goldstone bosons and thus massless. In the Standard Model of Particle Physics, matter (in the broad sense used in this paper) is divided into bosonic force carriers and fermionic matter (in a narrower sense of matter).\footnote{We are excluding the Higgs boson here, since, although it is a boson, it is not associated with one of the four fundamental forces.} But there is no reason why matter, in this narrow sense, could not also be bosonic, and it is not uncommon for new dark matter theories to postulate a bosonic dark matter field. If the associated particles self-interact (repulsively), they can form a superfluid Bose-Einstein condensate (BEC), which carries phonons. In other words, in this phase one cannot associate with the field a set of individual (nearly) collisionless particles; it is best described in terms of collective excitations. If these phonons cohere they can mediate a long-range force. If the phonons are described by the appropriate Lagrangian, they mediate a MONDian force. In order for the superfluid BEC phase to obtain, the De-Broglie wavelength of the $\Phi$-particles needs to be larger than the mean interparticle separation \cite[p.5]{berezhiani2016} \cite[\S 2.2]{annett2004} (or equivalently, the associated temperature needs to be below a critical value). In order for the phonons to cohere the BEC needs to be allowed sufficient time to thermalise \cite[p.5-6]{berezhiani2016}. If parameters are chosen such that those two conditions obtain only in galaxies, but not at larger scales, this would provide a natural reason for why the MONDian behaviour of the single novel entity appears only in galaxies. This approach to resolving the dark phenomenon stalemate is created and developed by Berezhiani and Khoury, under the name `Superfluid Dark Matter Theory'. We will outline the original version of their theory \cite{berezhiani2015,berezhiani2016} in this section.

%The theory takes General Relativity with a cosmological constant and `normal' matter content as its starting point, i.e. the Lagrangian is \todo{This is the incorrect Lagrangian. The whole point is that the matter Lagrangian is a function of the physical metric, not of the Einstein metric. This is the only and crucial fact that will determine our 'evaluation of Phi according to the spacetime criteria'.}

%\todo{Berezhiani 2016 p17: a0 depends on velocity dispersion... on cosmological scales it will be different form the value that MOND people take it to be.}

This theory postulates a self-interacting, massive, complex scalar field $\Phi$ with global $U(1)$ symmetry, via the following term:
\begin{equation} \label{eqBerezhiani}
	\mathcal{L}_{\Phi} = - \frac{1}{2} \left( |\partial_{\mu} \Phi|^2 + m^2 |\Phi|^2 \right) - \frac{\Lambda^4}{6(\Lambda_c^2+|\Phi|^2)^6} \left( |\partial_{\mu}\Phi|^2 + m^2|\Phi|^2 \right)^3
\end{equation}
with $\Lambda_c$ and $\Lambda$ two constants introducing two mass/energy scales.\footnote{$\Lambda_c$ ensures that the theory admits a $\Phi=0$ vacuum \cite[p.15]{berezhiani2016}.} (The main justification for this choice of Lagrangian comes from reverse engineering: it is this Lagrangian that will ultimately reproduce MOND in the Galactic regime, as discussed below.) 

%\todo{ Berezhiani 2017 p5: the interaction term makes means that we have a pseudo-superfluid, and the phonons will in fact acquire a mass via radiative corrections, even though this has no observable effects on galactic scales.}
Spontaneous symmetry breaking of the global $U(1)$ symmetry yields, in the non-relativistic regime, the following Lagrangian for the associated Goldstone bosons---these massless phonons being represented by the scalar field $\theta$, the phase of $\Phi = \rho e^{i(\theta + mt)}$:

\begin{equation} \label{eqX}
	\mathcal{L}_{T=0,\neg rel,\theta} = \frac{2\Lambda(2m)^{3/2}}{3}X\sqrt{|X|},
\end{equation}
with 
\begin{equation}  \label{eqX2}
	 X \equiv \dot{\theta} - m\Phi - \frac{(\vec{\nabla}\theta)^2}{2m},
\end{equation} 
where $\Phi$ is now interpreted as the external gravitational potential. (This identification receives its justification when one derives this non-relativistic description from the relativistic description, i.e.\ \hyperref[crucial]{Eq.(\ref{crucial})}, with $\tilde{g} = \tilde{g}^{SFDM}$ as given by \hyperref[gFDM]{Eq.(\ref{gFDM})}.) To lowest order in the derivatives, superfluid phonons are in general described  by a scalar field $\theta$ governed by a Lagrangian $\mathcal{L} = P(X)$, with $X$ given by \hyperref[eqX2]{Eq.(\ref{eqX2})} \cite[p.3]{berezhiani2016}\cite{son2002}. The specific choice of $P$ given by \hyperref[eqX]{Eq.(\ref{eqX})} uniquely determines the specific type of superfluid, namely one that interacts primarily through three-body interactions, i.e.\ with an equation of state $P \propto \rho^3$.

The interaction between the phonons and the regular (i.e.\ luminous) matter fields is then added to this Lagrangian as an ``empirical term'' \cite[p.8]{berezhiani2016} (as opposed to being derived from an interaction term in the fundamental Lagrangian). In the relativistic regime we may describe this via the metric $\tilde{g}_{\mu\nu}$, sometimes referred to as the physical metric---why such metrics deserve this name is discussed in \cite[\S 5.2]{martenslehmkuhl1}---because all luminous-matter fields couple universally to this metric (and to the Einstein metric $g_{\mu\nu}$ only indirectly, via this physical metric):\footnote{\label{newFDM} In a later paper \cite{berezhiani2017}, photons are not coupled to the effective metric, but to the Einstein metric. Some conclusions reached in section 5.2 of the prequel paper \cite{martenslehmkuhl1} do not apply to that alternative version of the SFDM theory.}
\begin{equation} \label{crucial}
\mathcal{L}_{\ rel,\ int} = \mathcal{L}(\tilde{g}_{\mu\nu},\psi^{\alpha},\psi^{\alpha}_{;\mu | \tilde{g}})
\end{equation}
with $\psi^{\alpha}$ the luminous-matter fields, and $\psi^{\alpha}_{;\mu | \tilde{g}}$ denoting the covariant derivative with respect to $\tilde{g}$. The physical metric of SFDM is inspired by that of TeVeS. 

TeVeS, a theory usually referred to as a modification of gravity, postulates two new dynamical fields---a real scalar field $\phi$ and a 4-vector field $A_{\mu}$---which, together with the Einstein metric $g_{\mu\nu}$, constitute the effective metric $\tilde{g}_{\mu\nu}^{TVS}$. This physical metric is disformally related to the Einstein metric \cite{bekenstein2004}\cite[\S 3.4]{clifton2012}.  That is, it does not stretch the Einstein metric equally in all directions, which would leave all angles and thus shapes invariant, as with a conformal transformation, but instead \emph{stretches} it by a factor of $e^{-2\phi}$ in the directions orthogonal to $A^{\mu} \equiv g^{\mu\nu}A_{\nu}$ while \emph{shrinking} it by the same factor in the direction parallel to $A^{\mu}$, where the Sanders 4-vector field $A_{\mu}$ is unit-time-like with respect to the Einstein metric ($g^{\mu\nu}A_{\mu}A_{\nu} = -1$) \cite{bekenstein1993,bekenstein2004,clifton2012}:
\begin{equation} \label{tevesmetric}
\begin{array}{rcl}
\tilde{g}^{TVS}_{\mu\nu} & = & e^{-\frac{2\alpha\Lambda}{M_{Pl}}\phi}(g_{\mu\nu}+A_{\mu}A_{\nu}) - e^{\frac{2\alpha\Lambda}{M_{Pl}}\phi}A_{\mu}A_{\nu} \\
& = & e^{-\frac{2\alpha\Lambda}{M_{Pl}}\phi}g_{\mu\nu}-2A_{\mu}A_{\nu}\text{sinh}(\frac{2\alpha\Lambda}{M_{Pl}}\phi) \\
& \approx & g_{\mu\nu} - \frac{2\alpha\Lambda}{M_{Pl}}\phi(g_{\mu\nu}+2A_{\mu}A_{\nu}),
\end{array}
\end{equation}
with $M_{Pl}$ the Planck mass and $\alpha$ a dimensionless coupling constant.

SFDM modifies TeVeS in two ways: one semantic and one syntactic modification. The semantic revision is to not add yet another (vector) field, but to identify the four-vector $A_{\mu}$ with the unit four-velocity $u_{\mu}$ of the previously added scalar field $\Phi$.\footnote{Of the normal fluid component; see below.} Syntactically, the TeVeS factor of 2 is generalised to obtain: 
\begin{equation} \label{gFDM}
	\tilde{g}_{\mu \nu}^{SFDM} \approx g_{\mu \nu} - \frac{2\alpha\Lambda}{M_{Pl}}\theta \left( \gamma g_{\mu \nu} + (1+\gamma)u_{\mu}u_{\nu} \right),
\end{equation} 
with TeVeS being recovered for $\gamma = 1$ (and a metric conformally related to $g_{\mu \nu}$ for $\gamma = -1$). 

The full relativistic form may be used to calculate gravitational lensing effects. 
%\todo{This doesnt' seem true anymore in berezhiani2017. See handwritten notes at the end of october 2018 on important points in berezhiani 2016 and 2017.} 
In the context of recovering the galaxy rotation curve component of the dark phenomena we only require the non-relativistic approximation:
\begin{equation} \label{eqTeVeSlike}
	\mathcal{L}_{\neg rel,int} = - \alpha \frac{\Lambda}{M_{Pl}}\theta \rho_b,
\end{equation}
where $\rho_b$ is the baryon density. The total effective, non-relativistic Lagrangian can then be shown to reproduce \cite[p.10-11]{berezhiani2016}, under suitable approximations (such as $\theta$ being static and spherically symmetric), the MONDian result $a_{MOND} = \sqrt{a_0 \frac{GMb(r)}{r^2}}$ for a baryonic particle with mass $M_b$, after identification of $a_0 = \frac{\alpha^3\Lambda^2}{M_{Pl}}$.

To a particle physicist, the fractional power of $X$ (\hyperref[eqX]{Eq.\ref{eqX}}), 3/2, although required if one aims to eventually regain MONDian behaviour \cite[\S IV]{berezhiani2016}, might seem strange---it is, for instance, less straightforward to draw corresponding Feynman diagrams. In condensed matter theory, such powers are far from rare. As mentioned, this specific fractional power corresponds to a phonon superfluid, with equation of state $P \propto \rho^3$.

Superfluidity only occurs at sufficiently low temperature. This naturally distinguishes between galaxies and galaxy clusters. Due to the smaller velocities in galaxies,\footnote{Given a mass $m$ and density $\rho$ \cite[Eqs.\ 8 \& 80]{berezhiani2016}.} the superfluid description is appropriate there,\footnote{Since the local phonon gradient induced by the Sun is too large to satisfy the criteria for a superfluid Bose-Einstein condensate, the condensate loses its coherence, which allows SFDM to avoid solar system constraints \cite[\S V]{berezhiani2016}.} exactly where MOND is successful. In the clusters the velocity/ temperature is too high, and one finds either a mixture of the superfluid and normal phase, or only the normal phase, suggesting that the theory might exemplify the usual successes of dark matter theories at that level.

Unfortunately it is non-trivial to piece together a complete, general (i.e.~valid at all temperatures) theory from the separate pieces provided above. Assuming that we would be able to find a relativistic generalisation of the free phonon Lagrangian (\hyperref[eqX]{Eq.\ref{eqX}}), we take it that Berezhiani and Khoury intend the following (relativistic) Lagrangian to correspond to a complete description of the superfluid regime, that is below $T_c$---or strictly speaking only at $T = 0$, since at finite temperatures correction terms would need to be added \cite{berezhiani2016}:
\begin{equation} \label{fullTzero}
	\mathcal{L}_{T = 0} = \mathcal{L}_{Einstein-Hilbert}(\g) + \mathcal{L}_{\theta}(\theta,\g) + \mathcal{L}_{luminous}(\tilde{g}_{\mu\nu}^{SFDM},\psi^{\alpha},\psi^{\alpha}_{;\mu | \tilde{g}^{SFDM}}).
\end{equation}
It is then tempting to postulate as the general Lagrangian:
\begin{equation} \label{tempting}
\mathcal{L}_{gen} = \mathcal{L}_{Einstein-Hilbert}(\g) + \mathcal{L}_{\Phi}(\Phi,\g) + \mathcal{L}_{luminous}(\tilde{g}_{\mu\nu}^{SFDM},\psi^{\alpha},\psi^{\alpha}_{;\mu | \tilde{g}^{SFDM}}).
\end{equation}
In order for this to reproduce the successes of dark matter theories in galaxy clusters and at the cosmological level, we would like the theory to take on or approximate  $\mathcal{L}_{Einstein-Hilbert}(\g) + \mathcal{L}_{DM}(\text{DM-field},\g) + \mathcal{L}_{luminous}(\g,\psi^{\alpha},\psi^{\alpha}_{;\mu | \g})$ in that regime (i.e.\ above $T_c$). For that to be the case, $\tilde{g}^{SFDM}$ needs to reduce to the Einstein metric in that regime. \emph{Prima facie}, this seems to work out nicely: $\tilde{g}^{SFDM}$ is given by the Einstein metric plus a phonon modification term which vanishes when $\theta$ is zero. Above the critical temperature there are indeed no phonons. However, although $\theta$ may indeed represent the phonon field below $T_c$, it is, in general, the phase of $\Phi$. As such there is no reason why it would, generally speaking, be zero or even very small at large temperatures. It is thus an open question what the full SFDM theory is.\todo{IGNORE: Lasha and Justin mention some possible relations between the T0 description and the full theory} 

%It is not totally clear from the original FDM papers how the above Lagrangians are to be pieced together. ............................................... Just take the Phi lagrangian plus the effective metric lagrangian. Interpreting theta as phonons (with u its associated unit four-velocity) would mean that the effective metric reduces to the Einstein metric when there are no phonons around. In that case we would retrieve GR plus an extra scalar field, making it plausible that the success of lambdaCDM are reproduced in that scenario. However, technically speaking theta is defined via ....... and hence can be nonzero even when the temperature is not below the critical temperature. Do we need to make the physical stipulation that the L-effectivemetric only applies when we are below the critical temperature? But how does that follow from the mathematics? 

%......................(To what Lagrangian this new term is added is controversial. The obvious suggestion would be the Einstein-Hilbert Lagrangian plus some Lagrangian for luminous matter, but we will discuss in \hyperref[discussion]{\S \ref{discussion}} that this cannot be correct, or at least not the complete story.)

\section{The field $\Phi$ matters and spacetimes} \label{spacetime-matter}

% XXX Could make this part of the previous section, but it definitely should not be part of the Jesus section

\todo{IGNORE: In section 4.2 of Patrick's new JBD paper he argues against mass as necessary for being matter---which coheres with the reviewer who thinks that this is too strong a criterion.  Ask him permission to cite.}

In the companion paper, we introduced a family of criteria for something to be matter \cite{martenslehmkuhl1}. The strongest such criterion was `having mass'.\footnote{More precisely: The object under consideration changes in a regular fashion. This change is (partially) in response to something external, and the interaction describable in terms of coupled differential equations, in such a way that the object can be said to carry exchangable energy (and momentum) representable by a stress-energy tensor, $T_{\mu\nu}$, part of which is due to the rest mass of the object. In other words: the object is massive.} It also introduced a set of jointly sufficient criteria for allowing something to be interpreted as spacetime structure: a) being an object that is faithfully representable by a differentiable manifold with an affine connection and a Lorentzian conformal structure (i.e.\ an equivalence class of Lorentzian metrics) such that b) there is a well-defined initial value problem, and such that c) if test particles and/or light rays were around they would, for any choice of initial conditions, follow the timelike and null geodesics, respectively, of the same family of geodesics (i.e.\ the (strong) geodesic criterion), and such that d) special relativity is locally valid, in the sense that no non-negligible curvature terms appear in the equations of motion (i.e.\ the chronogeometricity criterion).

We argued that, according to these criteria, it is clear that the scalar field $\Phi$ occuring in SFDM is as much `matter' as it could possibly be: it fulfils the strongest matter citerion of having rest mass, and all the prior criteria that the mass criterion builds on. Denying that $\Phi$ is (dark) matter would imply that even paradigmatic instances of matter such as electrons (as they appear in the Standard Model of particle physics) are not matter either. 

However, the $\Phi$-field arguably scores just as well in terms of being (an aspect of) spacetime as it does in terms of being matter. We argued that, in the superfluid regime, $\tilde{g}_{\mu\nu}$ satisfies the above criteria for something to be interpreted as (an aspect of) spacetime. Given that $\tilde{g}_{\mu\nu}$ is defined in terms of $\Phi$, one could argue that $\Phi$ is thus just as much (an aspect of) spacetime as it is matter. The rest of the paper is devoted to introducing and evaluating three groups of viable interpretations of $\Phi$ as it features within SFDM, in light of the above, with the aim of providing an interpretational chart that can serve as a tool for describing and ultimately understanding (theories with) fields that are both (dark) matter and (aspects of) spacetime.

\section{Interpretations by analogy with theology} \label{jesus}

%\todo{Use Consubstantiality interpretation instead of consubstantiality interpretation. Use `doctrine' instead of `mythology'. Mention clearly that the constubstantiality interpretation is named BY ANALOGY with Trinitarianism within Christian theology/doctrine about Jesus. }

%\todo{My 3 questions about paper two. 
%Doctrine instead of mythology. 
%Demi god is a problem. 
%Jekyl Hyde include trafo (Dennis: impostor, but without pretending. Ik: conventionalism? Or related to the two phases oh phi. But in superfluid phase it plays both roles. Hulk and Prof hulk is better. Edit: no, Dennis must have meant She-hulk, who (mostly) remains her personality when in Hulk form.) decision leave it at Jesus. Re Dennis on Einstein on fields, maybe cite the discussion of Laue at the end of the Equivalence principles paper by Dennis.
%Superman Clark Kent as impostor interpretation.
%Voldemort and Quirrell, and Janus. 
%Dennis said something about personality of one person and look of another person.... Maybe hulk and Prof hulk, or she hulk. 
%Jekyll and Hyde would apply if only one role went with each phase. Same for Hulk. Different for She-Hulk.
%}

As we have seen in the previous section, the $\Phi$-field in SFDM scores just as high in terms of being matter as it scores in terms of being (an aspect of) spacetime. In the following sections we will see interpretations that claim that it is fundamentally one and merely emergently the other, or even that what we see is a breakdown of the very distinction between spacetime and matter. 

Both of these (families of) interpretations are rather radical if one is used to the classical distinction between spacetime and matter. And a conservatively minded philosopher/physicist might well ask: are we not overreacting to the existence of $\Phi$ with either option?

By analogy, let us consider corresponding scenarios from theology where we may ask what a new type of entity meant for the distinction between human and god, which arguably was challenged just as much---or just as little---by this new entity as the distinction between spacetime and matter is challenged by $\Phi$. 

Consider first mythologies with demigods, like Heracles in Greek mythology. A demigod has two parents, one a god and one a human, and a demigod was thus 50\% human and 50\% god. Zeus (Heracles' father) was 100\% god, and Alcmene (Heracles' mother) was 100\% human. Contrast this with the Christian Jesus, understood according to the consubstantiality thesis or Trinitarianism:\footnote{See, for instance, Catechism 467.} Jesus is not a demi-god, nor indeed either purely human or purely god; he is supposed to be 100\% god and 100\% human, even while on Earth. Puzzling about how this could be is one of the great mysteries of catholic scholarship. But despite all this, it would not be sensible to say that this concept of Jesus brought about the breakdown of the very distinction between god and human. Jesus is, in a conceptual sense, an anomaly, the single being that is seen as both fully human and fully god. However, this does not mean that any human (even a faithful catholic) should have any problem with classifying themselves and their neighbours as clearly human and non-god. 

How is this analogous to the physical fields that we are attempting to interpret? Recall first, from the prequel paper \cite{martenslehmkuhl1}, that non-trivial metric solutions $\g$ in GR satisfy up to matter criterion D (action-reaction) or E (energy definable in region). This suggests that $\g$ is not 0\% matter, but, if one requires a stronger matter criterion as a sufficient criterion for being matter, then $\g$ is not 100\% matter either. Combining this with the claim that $\g$ is 100\% spacetime (modulo the caveats mentioned in the prequel paper) suggests that the analogue of non-trivial metrics in GR lies somewhere between Heracles and Jesus. When it comes to $\Phi$, being fully (dark) matter and (an aspect of) spacetime, there is no mistaking its analogue: Jesus.\footnote{Other similar analogues that one might wish to consider are the two-faced gods Janus from Roman mythology and Culsans from Etruscan mythology, although the possibility of having a second face is not as conceptually surprising and controversial as one might think it is to be both spacetime and matter or a human and a god. Perhaps the double personality of Voldemort-Quirrell is a somewhat better analogy in this regard. If one were to consider a field which plays only the spacetime role when it is in one phase and only the matter role when it is another phase---as Berezhiani and Khoury consider to be the case for $\Phi$ (\hyperref[fundamentalist]{Section~\ref{fundamentalist}})---then the transformations between Bruce Banner and the Hulk, and between Dr.\ Jekyll and Mr.\ Hyde may be good analogies. If instead---as we believe to be the case for $\Phi$ (\hyperref[fundamentalist]{Section~\ref{fundamentalist}})---a field is always and everywhere matter but only in one specific phase also spacetime, the comparison with Jennifer Walters and She-Hulk seems more appropriate, as the personality of Walters is still (mostly) there even when in She-Hulk form. Finally, Hossenfelder \cite{hossenfelder2017} uses `imposter' terminology, which may suggest a Clark Kent/Superman analogy. However, this would not suggest an interpretation equivalent to the consubstantiality interpretation, but something more in the spirit of the elitist interpretations (\hyperref[fundamentalist]{Section~\ref{fundamentalist}}) (but see fn.\ref{imposter}).} Hence, we call the interpretation of $\Phi$ discussed in this section, by analogy, the consubstantiality interpretation. A similar interpretation for objects that are only partially matter and partially spacetime would be called the demi interpretation, by analogy with demigods such as Heracles.

Advocates of the two other, revisionary groups of interpretations discussed in the following sections may or may not grant that the existence of `demigod objects' (such as Heracles and $\g$) leaves the associated distinctions (god--human or matter--spacetime) untouched, but they insist that `full hybrid objects' (such as Jesus or $\Phi$) are either problematic for the associated distinctions or at least require us to say something more. This is one of the ways in which considering the issue of distinguishing matter and spacetime in the SFDM context goes beyond considering it within GR. On the consubstantiality interpretation however, the existence of a field like $\Phi$ that is just as much matter, namely maximally, as it is an aspect of spacetime structure, does not cause any big problems for the spacetime/matter distinction. The senses in which $\Phi$ is matter and spacetime are on a par. Nothing further needs to be said. Sure, $\Phi$ is rather unusal and weird in this way, but as long as it is in a clear minority with respect to other entities, the spacetime/matter distinction stands strong: it remains coherent, applicable, objective and useful. In a strange way, one might even say that the analysis of $\Phi$ has shown us what it really means to be matter, just as the concept of Jesus, some would say, has shown us what it is to be human.

The consubstantiality interpretation is a viable interpretation of $\Phi$. To the extent that there might be an objection to this interpretation it would be that one or both of the other two groups of options on the table, discussed below, provide a better or more complete account. Both groups of alternatives insist that there is more to be said about $\Phi$ being 100\% matter and 100\% spacetime. Hence, we will compare and contrast the consubstantiality interpretation with these other interpretations in \hyperref[cartography]{Section~\ref{cartography}}, after the latter have been introduced and analysed.

\section{Fundamentalist interpretations} \label{fundamentalist}

\begin{table}
	\begin{tabular}{cc||c|c}
		& & \multicolumn{2}{c}{MG}  \\ \cline{3-4}
		& & fundamental & emergent \\ \hline \hline
		\multirow{4}{*}{DM}	& \multicolumn{1}{|l||}{fundamental} & {\bf \emph{(MG+DM)-fundamentalism}} & {\bf \uline{DM-fundamentalism}} \\ 
				& \multicolumn{1}{|l||}{} & \emph{(or: dualism)}  &					\\ \cline{2-4}
		& \multicolumn{1}{|l||}{emergent} &	{\bf \uline{MG-fundamentalism}}	& {\bf \emph{non-fundamentalism}} \\
		& \multicolumn{1}{|l||}{} & & \emph{(or: neutral monism)}
	\end{tabular}
	\caption{\label{fundtable}Matrix of views that for each of the DM and MG aspects distinguish between them being fundamental or merely emergent aspects. Egalitarian views are in italics; elitist views are underlined.}
	\label{matrix}
\end{table}

%\cniels{Dennis, you suggested naming the non-fundamentalist approach something Greek, namely the name that belongs to the view that distinguishes matter and form. The word is `hylomorphism', but since `hule' is often translated as matter rather than substance, this could be very confusing in our case (since what we want to say is that their is one neutral substance, which could take on two different forms (namely matter or spacetime/gravity)).}

The fundamentalist interpretations go beyond the consubstantiality interpretation by claiming, for each of the DM and MG aspects, whether $\Phi$ exhibits them fundamentally or not. This results in a 2 by 2 matrix of logically possible views, see \hyperref[matrix]{Table~\ref{matrix}}. On the (MG+DM)-fundamentalist interpretation, both aspects are fundamental aspects of $\Phi$. On the non-fundamentalist interpretation $\Phi$ is fundamentally neutral, i.e.\ neither DM nor MG; both of these aspects are  non-fundamental, derivative, or even emergent. Besides these two egalitarian positions\footnote{Not to be mistaken with the notion of `egalitarian interpretation' in \cite{lehmkuhl2008}.} there are two elitist positions, on which only one of the DM and MG categories is a fundamental aspect or property of $\Phi$, with the other category being merely a derivative or emergent aspect or property---as opposed to the divinity and humanity of Jesus being, in all relevant senses, on a par. Therefore the egalitarian positions can be considered as being closer to (the spirit of) the consubstantiality interpretation. Similarly, the non-fundamentalist interpretation is closest (in spirit) to the breakdown interpretations (see below), as both aspects being metaphysically non-fundamental opens the door to them not being conceptually fundamental either. (The relations between these interpretations will be discussed further in \hyperref[cartography]{Section~\ref{cartography}}.)

%On the fundamentalist or impersonator interpretations, only one of the DM and MG categories is a fundamental aspect or property of the object under consideration, with the other category being merely an emergent aspect or property---as opposed to the divinity and humanity of Jesus being, in all relevant senses, on a par.

Within the category of elitist interpretations, it may seem most plausible that $\Phi$ is fundamentally dark matter and emergently a modification of gravity and spacetime (DM-fundamentalism), rather than vice versa (MG-fundamentalism). The MG-nature is important only in galaxies. Hence, $\Phi$ merely plays the role of a modification of spacetime in that regime; \todo{IGNORE: Write footnote: We mean emergence in this sense, not in the sense of whether certain symbols do or do not appear in the fundamental lgrangian. For instance, TeVeS indeed does not mention g-effective as a separate symbol, but is is clear from that same `fundamental' formulation that matter does only feel that effective metric. The lesson to be learnt: don't let notation misguide you.} it is an MG-impersonator,\footnote{\label{imposter}Hossenfelder \cite{hossenfelder2017} uses `imposter' terminology, but we do not want to suggest that what is impersonated/mimicked is fictive; it is real; it is just not fundamental. Compare this to a chair, which despite not being fundamental is still real rather than a fiction.} not fundamentally a modification of spacetime.

Arguably, this is what the creators of SFDM have in mind. The following textual evidence places them in the right column of \hyperref[matrix]{Table~\ref{matrix}}.
\begin{quote}
	In contrast with theories that propose to fundamentally modify Newtonian gravity, in this
	case the new long-range force mediated by phonons is an	\emph{emergent} property of the DM superfluid medium. (italics in the original) \cite[p.3]{berezhiani2017}
\end{quote}
\begin{quote}
	Unlike most attempts to modify gravity, there is no fundamental additional long-range force in the model. Instead the phonon-mediated force is an \emph{emergent} phenomenon which requires the coherence of the underlying superfluid substrate. (italics in the original) \cite[p.16]{berezhiani2017}
\end{quote}
The name given by them (but not by Hossenfelder, who calls it a modified gravity theory! \cite{hossenfelder2017}) to their theory, i.e.\ superfluid dark matter, further narrows it down to DM-fundamentalism. In the remainder of this subsection we will focus on evaluating this interpretation---it being further removed from the consubstantiality interpretation than the egalitarian interpretations and \emph{prima facie} more plausible than MG-fundamentalism. Comments regarding the other three interpretations will be made along the way.

Before proceeding it is important to note that Berezhiani and Khoury seem to operate under a different understanding of the dark matter aspect of $\Phi$. Whereas we take the property of `having mass' (or one of the weaker matter criteria \cite{martenslehmkuhl1}) to be what makes $\Phi$ dark matter, they seem at places to identify the DM nature of $\Phi$ with its particle nature, that is the phase in which $\Phi$ is appropriately described as individual collisionless particles, i.e.\ the phase above $T_c$ (as opposed to the MG-phase below $T_c$ where a description in terms of collective excitations---phonons---is considered more appropriate). In their words: ``DM and MOND components have a common origin, representing different phases of a single underlying substance'' \cite[p.3]{berezhiani2016}. It turns out, ironically, that not only does their understanding of the DM nature of $\Phi$ undermine the main argument in favour of DM-fundamentalism more obviously than our criterion does (\hyperref[DMfundarg1]{\S \ref{DMfundarg1}}), it also comes with its own strong argument against DM-fundamentalism (\hyperref[DMfundarg2]{\S \ref{DMfundarg2}}). After a brief discussion of these two arguments follows a more lengthy discussion of a general group of problems that DM-fundamentalism faces on either understanding of the DM nature of $\Phi$. 

\subsection{The scope argument} \label{DMfundarg1}

As briefly hinted at before, one line of argument in favour of DM-fundamentalism is the following. $\Phi$ exhibits its DM nature everywhere in the universe; this is not the case for its MG nature (which appears predominantly in galaxies). Hence, the DM nature of $\Phi$ is more fundamental than the MG nature. Let us call this the scope argument.\footnote{Thanks to Katie Robertson and Alex Franklin for pushing us on this line of argument.} It is not immediately obvious why this conclusion would follow. There are at least three ways of fleshing out this argument. Two flawed ways will be discussed here. A better version is touched upon in \hyperref[DMfundarg3]{\S \ref{DMfundarg3}}.

Before evaluating the validity of this argument, it is important to note that it is not even sound to begin with if one understands, with Berezhiani and Khoury, the DM and MG natures of $\Phi$ as each being a distinct phase of $\Phi$. It is of course not impossible for several phases of a substance to coexist---at the triple point of water there are even three phases that co-exist---but we would typically expect a single phase to be dominant. At some regions and times this will be the DM phase, at others the MG phase. The premises of the above argument do not hold. They do seem to hold\footnote{However, if the final theory of SFDM is indeed such that the modification of the Einstein metric at $T \gg T_c$ goes to zero---see the end of \hyperref[FDMintro]{Section~\ref{FDMintro}}---then it seems to be the case that the MG aspect is instantiated everywhere after all, albeit (partially) trivially.} on our preferred understanding of the DM-nature of $\Phi$, i.e.\ its massiveness: $\Phi$ is always and everywhere massive, but does not always and everywhere form a superfluid BEC. (It should of course be noted though that there is not yet a complete theory, exactly because there is no justification for the $\theta$-modification to the Einstein metric vanishing outside of galaxies.)

A first way of cashing out the scope argument starts from the claim that because of the universal scope of the DM aspect of $\Phi$ and the non-universal scope of the MG aspect, the former is more important than the latter. Hence, $\Phi$ is DM in a `more real', a `more fundamental' sense than that it is MG. However, not only is `importance' a vague, subjective, anthropocentric term, but this version of the argument conflates the colloquial meaning of fundamental---important---with the more philosophical, technical meaning that is intended in the formulation of DM-fundamentalism (see \hyperref[DMfundarg3]{Subsection~\ref{DMfundarg3}}).

A second way of cashing out the argument phrases the non-universality of the MG-aspect of $\Phi$ as that aspect `appearing' in galaxies and `disappearing' outside of galaxies. On an understanding of `to emerge' as this notion of `to appear', DM-fundamentalism follows. But this similarly mistakes the relevant technical notion of emergence with the colloquial meaning. That a property does not appear always and everywhere does not imply that it is non-fundamental. Consider electromagnetism. The magnetic field inside an infinitely long solenoid will be non-zero, but outside the solenoid it will be zero. We do not usually take this to imply that the magnetic field is non-fundamental (within electromagnetism). Similarly, in electrostatics a Faraday cage may be used to shield off an external electric field. That the electric field disappears within the cage is not usually taken to imply that it is non-fundamental. This second version of the argument seems to mistake essentialism---for a property to be essential to an object it must be instantiated always and everywhere in all possible worlds inhabitated by that object---for fundamentalism.

%Especially this second version of the argument suggests that 

%viscosity disappears (below a certain velocity) in superfluids and thus in galaxies. 

%Unclear that it is zero outside galaxies, as theory is incomplete

What perhaps inspires these two versions of the argument is that paradigmatic emergent theories, such as Newtonian gravity as compared to General relativity, are not universally valid but only (approximately) valid in certain limiting regimes (see the `formal asymmetry criterion' in \hyperref[DMfundarg3]{Subsection~\ref{DMfundarg3}}). It is indeed the case that emergent theories that are the limit of a more fundamental theory therefore have a limited scope of applicability. But it does not follow that the implication goes in the other direction: from a non-universal scope of some property it does not follow that that property can only be described by an emergent theory that is the limit of a more fundamental theory.

On colloquial understandings of the notions `fundamental' and `emergent' the scope argument is not valid. For it to stand any chance it must connect the non-universality of the MG nature of $\Phi$ to the technical notions of `fundamental' and `emergent' that are intended in the formulation of DM-fundamentalism. These technical notions will be outlined in \hyperref[DMfundarg3]{\S \ref{DMfundarg3}}, where the scope argument will be touched upon when the notion of `ontological dependence' is discussed.

\subsection{Water analogy} \label{DMfundarg2}

Remember that the superfluid BEC, which carries the phonons that mediate the MONDian force, is a phase of $\Phi$; this effect appears at temperatures below the superfluid phase transition. If one follows Berezhiani and Khoury in taking the DM-aspect to be the phase of $\Phi$ above the critical temperature (but not on our understanding of DM), DM-fundamentalism starts to sound a lot like saying that water vapour is more fundamental than liquid water,\footnote{In the phase diagram of water, one can even move between the liquid and vapour phases without crossing any phase transition, namely by circumventing the critical point. The only difference between these two `phases' is then their density (which would justify a fundamental--emergent distinction even less). (Thanks to Stephen Blundell for this point.) The two phases of $\Phi$ are thus perhaps better compared to ice and liquid water.} or liquid water more fundamental than ice. We do not tend to think of the different phases of water in that way;\footnote{At least some phase transitions (such as the transition of $\Phi$ at $T_c$) are associated with a breaking of the symmetries of the system. From a particle physics perspective, e.g.\ in the context of grand unified theories, it may seem common to associate with this change in symmetry a difference in fundamentality. However, this notion of fundamentality in terms of unification or scope, if indeed a relevant notion of fundamentality at all, differs from the concepts of fundamentality and emergence as intended in this paper (\hyperref[DMfundarg3]{\S \ref{DMfundarg3}}).} they are just different but equal states of being of a single underlying entity. (Moreover, if we would encounter one phase only or mainly in galaxies, and the other phase all over the universe, this would not change anything.) This resonates with egalitarian interpretations of the different phases of water.\footnote{In \hyperref[DMfundarg3]{\S \ref{DMfundarg3}} it will be pointed out that in galaxies, generally speaking, a dynamical mixture occurs: both the DM and the MG aspects simultaneously contribute to the total dynamics of luminous matter. Does this consistute a (relevant) disanalogy with water? As pointed out before, even water can co-exist in multiple phases; at the triple point of water there are even three phases that co-exist. However, generally speaking we would expect there to be a single (dominant) phase, which would indeed speak against the possibility of there generally being a dynamical mixture of MG and DM if both of these are understood as distinct phases of $\Phi$. The tension resolves when we note that the `DM' contribution arises from the mass of $\Phi$, which should not have been labeled as a `DM' contribution if one defines the DM aspect of $\Phi$ as its particle phase. The only oddity that remains is that Berezhiani \emph{et al.} do in fact label that contribution as the DM contribution \cite[Eq.\ 47]{berezhiani2017}; we of course agree with this labeling but it is inconsistent with their definition of DM.} On a strong reading of `underlying' in the earlier quote by Berezhiani and Khoury---``DM and MOND components have a common origin, representing different phases of a single underlying substance'' \cite[p.3]{berezhiani2016}---non-fundamentalism would be the favoured egalitarian interpretation. At the fundamental level there is just a neutral entity, water, H$_2$O ($\Phi$), which can appear in different forms---ice, liquid, vapour (phonons, individual collisionless particles)---without any single one of these phases being essential to water ($\Phi$), or a fundamental aspect of it. 

To sum up: if one were to equate DM to the phase in which an particle description is appropriate---which Berezhiani and Khoury do but we do not---and MG to the phase in which a phonon description is appropriate, DM-fundamentalism would be problematic.

\subsection{Fundamentality \& Emergence} \label{DMfundarg3}

%Secondly, the above formulation of the argument for DM-fundamentalism confuses the colloquial meaning of fundamental, i.e.\ important, with the more technical, metaphysical notion that is at stake here. Once we spell out what is meant by `fundamental' and `emergent', we will see that the DM-fundamentalist interpretation faces another group of interrelated objections.

It is about time that we spell out in a bit more detail what one might mean by `fundamental' and `emergent' in the context of DM-fundamentalism (and the other three fundamentalist interpretations)---other than the too colloquial meanings of `important' and `appearance', respectively, as discussed in \hyperref[DMfundarg1]{\S \ref{DMfundarg1}}. Consider the variant of the notion of emergence that takes it to be a relation between descriptions, i.e.\ between different dynamical stories that could be told to account for the observations. If the DM-fundamentalist interpretation is appropriate, we might expect (at least) the following picture: 
\begin{enumerate}
	\item {\bf Metaphysical asymmetry:} the emergent description (i.e.\ MG) ontologically depends on the fundamental description (i.e.\ DM) \cite{lehmkuhl2018};
	\item {\bf Formal asymmetry:} the emergent description is obtained from the fundamental description via an irreversible mathematical operation \cite{knox2016};
	\item {\bf Scale separation/asymmetry:} the fundamental description is a description of smaller length scales and higher energies; the emergent description is a description of larger length scales and lower energies.
	\item {\bf Dynamical separation/autonomy:} the fundamental description is a description in terms of fundamental concepts (i.e.~DM) only, and, ideally, the emergent description is a description in terms of emergent concepts (i.e.~MG) only. 
\end{enumerate}

These criteria may be illustrated by comparing, say, an elephant as described by zoology with its description in terms of the underlying, more fundamental Standard Model of particle physics. There could be quarks without there being elephants, but not \emph{vice versa}. The emergent elephant featuring in Zoology is a coarse-grained version of the underlying description; from it one cannot retrieve the full state of the quarks. Zoology is a low-energy/ large distance limit of the Standard Model of particle physics. Zoologists do not need to talk about quarks; particle physicists do not need to refer to elephants.

We will discuss each of these components in turn. Although the metaphysical symmetry condition is satisfied in some sense in SFDM, there is strong reason to doubt that the formal asymmetry and dynamical separation apply. The scale asymmetry definitely does not apply.

%2.1

\subsubsection{Metaphysical \& formal asymmetry}

Is there a metaphysical and/or formal asymmetry between the two descriptions in virtue of which we may call the DM description the fundamental one and the MG description the emergent one? Start with the formal asymmetry. The literature is riddled with examples of such irreversible mathematical operations---idealisations, such as the thermodynamical limit, or abstractions, such as coarse-graining or summation \cite{knox2016,franklinknox2018}. Is the phonon description within SFDM arrived at via such an irreversible operation? 

\todo{IGNORE: Re the Franklin Knox point that phonons are reducible and there is this no ontological asymmetry so they might well be fundamental: this would imply that fundamental things can stop existing and come into existence over time. Are we ok with this? If not, perhaps we should take constitution as an appropriate asymmetric ontological relation. 
}

Franklin and Knox describe phonons as they appear in a crystal lattice of atoms \cite{franklinknox2018}. They convincingly argue that moving from the atomic displacement variables to a description in terms of phonon variables provides novel explanatory power. (Similarly, in SFDM, the phonon description explains the MONDian behaviour in galaxies.) They suggest that this explanatory novelty constitutes a relevant notion of emergence, differing from the standard notions associated with mathematically irreversible operations. In fact, the two descriptions are dual. Complete translations between the descriptions are possible in either direction. In light of this, they report David Wallace as pointing out that their notion of emergence fails to satisfy what we have called the metaphysical asymmetry criterion. They concede that a full defense of the emergent nature of phonons would require a further asymmetric relation between the two descriptions. 

This opens up an interesting possibility. What if there is no such asymmetric relation? Would both descriptions then be equally fundamental? Or, arguably, with the phonon description providing novel explanatory power, that description seems the more fundamental one. On the alternative definition of DM as the particle nature of an underlying field, this would suggest, in the context of SFDM, an inverted fundamentalist interpretation, with the MG aspect of $\Phi$ being fundamental and the DM aspect being merely emergent: the MG-fundamentalist interpretation. 

\todo{IGNORE: this option is only available on the particle criterion for DM. However, above the critical temperature the particle description is better.... so what has the most explanatory power/is the best description depends on phase/temperature.... this seems a lot like the non-fundamentalist interpretation: neutral field $\Phi$, with appropriate different descriptions at different temperatures.}

Is there then a further, metaphysical asymmetry that blocks the conclusion that phonons are fundamental? One may wish to consider several specific options. One specific option would be composition: phonons in an atomic crystal are `composed' of the collective behaviour of many many atoms. Another specific option would be a part-whole (i.e.\ mereological) asymmetry. Our preferred candidate for the relevant metaphysical asymmetry is more general, namely ontological dependence, which holds if and only if the phonons cannot exist without the atomic displacements but the atomic crystal could exist without there being phonons. One possible negative response is that in the SFDM context there is no underlying atomic crystal that could compose the phonons; it is much less clear that particle excitations of $\Phi$ have the same metaphysical robustness as atoms. A stronger negative response points out that, since the two descriptions are dual, it is not only the case that the phonon variables are linear functions of the atomic displacement variables, but the latter are also linear combinations of the former \cite[esp.~p.75]{franklinknox2018}. Duality is a symmetric relation; phonons are as much composed of or a part of atomic displacements as \emph{vice versa}. Now, classically we can indeed conceive of an atomic crystal at rest (i.e.\ all atomic displacements are 0 and remain 0), which would seem to imply that it is possible for an atomic crystal to exist without any phonons existing `in' it. It might be retorted that there would still be an ideal phonon, namely one of infinite wavelength and hence zero energy. A stronger response is that, regardless of phonons being ontologically dependent or not on a classical crystal, it is quantum mechanically not possible for a crystal to exist without there also being phonons. It is not dynamically possible for quantum mechanical constituents of a crystal to be at rest; the lowest energy state is not zero and does consist of a phonon (cf. the zero-point energy of a quantum harmonic oscillator). Phonons are thus not ontologically dependent on an atomic crystal. That being said, we will see below when discussing `dynamical separation' (and in particular the two-fluid model), that the SFDM case is much more complex than the atomic crystal that Franklin and Knox concern themselves with. Hence, a full evaluation of the metaphysical and formal asymmetry criteria is outside of the scope of this paper. However, the above discussion serves to show that one should by no means assume that a phonon, being a `collective excitation', \emph{by default} satisfies the metaphysical asymmetry criterion, let alone the formal asymmetry criterion. 

%Moreover, in the context of SFDM there is no underlying lattice description to begin with. In that context one might then suggest a difference in scope as the relevant asymmetry\footnote{Thanks to Katie Robertson for suggesting this.}: the phonon description covers only part of the universe. But this is just to reiterate the misguided argument that one phase of a substance is more or less fundamental than another. 

%\cniels{NIELS, EVEN IN THE ATOMIC CRYSTAL CASE, IT'S NOT CLEAR THAT THE ATOMIC DISPLACEMENTS ARE MORE FUNDAMENTAL IN THE SENSE OF THE PHONONS BEING ONTOLOGICALLY DEPENDENT ON THEM. This might seem the case, classically, where we could have all the atoms at rest without any phonons. But quantummechanically, the lowest energy state still has a phonon. It's not dynamically possible for the atom to be at rest... just as a quantum harmonic oscillator has a zero-point energy, or whatever it is called. Edit: hmmm, in the 2FM, at T0, there are no phonons. Or is there a phonon with an infinite wavelength and hence zero energy?}

Phonons are not ontologically dependent on the atomic crystal: although they indeed cannot exist without the crystal, it is not the case that the crystal could exist without there being phonons. If a crystal exists, then phonons will also exist. However, there is one obvious way to get rid of the phonons: melt the crystal. In that sense, even though a crystal ensures the existence of phonons, it is still possible for atoms to exist (albeit not in crystal form) without phonons existing. Phonons do not ontologically depend on an atomic crystal, but they do seem to ontologically depend on the atoms. Similarly, even if phonons are indeed not ontologically dependent on the superfluid BEC, one could `melt' this BEC by increasing the temperature, which makes the phonons disappear whilst leaving individual collisionless (massive) particles behind. The possible existence of particle excitations of $\Phi$ without there being phonons (which occurs typically in galaxy clusters and at the cosmological scale) then seems to suggest that phonons are ontologically dependent on the particles. This then finally suggests the most viable form of the scope argument: the universal occurrence of the DM-nature of $\Phi$ and the non-universal occurrence of its MG-nature suggests that the latter is ontologically dependent on the former, and since `ontological dependence' \emph{is} (part of) the relevant understanding of the fundamental-emergent dichotomy, DM-fundamentalism is favoured.

A first thing to note is that, even though in our universe it might well be the case that there are spacetime regions with particles but without phonons, it is still true (for finite $m$) that in each world dynamically allowed by SFDM, there will be a finite critical temperature below which $\Phi$ will be describable in terms of phonons such that the spacetime criteria are satisfied. In other words, the existence of $\Phi$ does guarantee that in each dynamically allowed world phonons could be made to exist as long as you (can) cool down the system sufficiently. 

To evaluate this final version of the scope argument, we need to again distinguish between our definition of DM as `massiveness' (or a weaker matter criterion \cite{martenslehmkuhl1}) and the alternative definition `being describable in terms of individual particles'. Start with the latter. One way to question the scope argument is as follows. Sure, the `melting story' above shows us that the particle excitations of $\Phi$ (DM) could exist without there being phonons (MG). But to prove ontological dependence of the latter on the former, it also needs to be the case that phonons could not exist without there being particles or atoms. In the case of an atomic crystal that is obviously the case. But in the case of a quantum field and especially a superfluid BEC one might well push back against the claim that there are, in any meaningful sense, individual particles. The literature on this is vast and we will not dwell on it here \cite{margenau1944,french1988,malament1996,halvorson2002,french2006,falkenburg2007,muller2008,muller2009,kuhlmann2010,ruetsche2011,caulton2013,huggett2014,muller2014,boge2018,brown2018,dorato2018,meynell2018,stoeltzner2018,wuethrich2018}. If it is correct that there are not in any meaningful sense individual particles in a superfluid BEC, it would be the case that phonons and particles are both ontologically independent of each other, and that neither of them exists everywhere in a SFDM-model that could describe our actual universe. 
%In that sense there is no temperature-independent statement about whether the MG aspect is ontologically dependent on particle excitations.

Consider now the scope argument on our (massiveness) definition of DM. First we may ask whether the MG-behaviour of $\Phi$ is ontologically dependent on its mass. The phonon-mediated force is clearly not reducible to a gravitational contribution of $\Phi$ to the dynamics of luminous matter: it arises from a direct coupling of the phonons to the baryons (Eqs.\ \ref{eqTeVeSlike} \& \ref{crucial}). It should be noted that the evolution of the phonons depends on $m$, as is clear from $\mathcal{L}_{\theta}$ (Eqs.\ \ref{eqX} \& \ref{eqX2}). But this is a form of dynamical dependence. We are here interested in whether the existence of the phonons (and the phonon-mediated force) depends on $m$, not their behaviour. The answer is negative, in the following sense. For a very small $m$, the De-Broglie wavelength and the critical temperature become very large. Hence, for a very small $m$, $\Phi$ is practically always and everywhere a superfluid BEC, with the associated phonons mediating the modification of gravity. 

The MG-aspect of $\Phi$ does not ontologically depend on its DM-aspect. Nevertheless, the MG-aspect may still depend on $\Phi$, and it does: $\Phi$ can exist without phonons, as it does, for instance, outside of galaxies in a model that is supposed to represent our actual world, whereas the phonons cannot exist without $\Phi$. In that sense, the MG description does ontologically depend on the $\Phi$ description, and since we have chosen $m$ to be finite, this $\Phi$ description is a DM description.

In conclusion: given the above mentioned lack of a complete underlying theory of SFDM, it is at this stage not possible to determine whether the phonon description is dual to the missing underlying description; but given Franklin \& Knox's discussion of phonons in an atomic crystal we should refrain from simply assuming that the formal asymmetry condition is satisfied by SFDM. Regarding the metaphysical asymmetry condition: phonons are not ontologically dependent on the superfluid BEC, nor on the mass of $\Phi$. However, it is possible for $\Phi$ to exist without carrying phonons, but not for phonons to exist without $\Phi$. For finite $m$, the phonons are thus ontologically dependent on the $\Phi$ description, which is a DM description (on our criteria for DM).

\subsubsection{Scale separation}

Next we consider the scale separation criterion. Although galaxies are in this context (i.e.~compared to galaxy clusters and the cosmological scale) indeed relatively small,\footnote{It should be noted that, at a yet smaller scale, namely that of individual stars, no condensation occurs and thus no MONDian behaviour emerges (which happens to be good news in light of solar system constraints \cite{berezhiani2016,berezhiani2017}).} they correspond to relatively low energies, as superfluidity occurs \emph{below} a critical temperature.\footnote{Note that not only are the scales inverted from what is required or expected, it is not even the case that there is a universal length scale. The second condition for the existence of a coherent BEC (\hyperref[FDMintro]{section \ref{FDMintro}}) requires sufficient time for thermalisation to occur, which will differ per galaxy.} One might retort: so be it---the concepts `fundamentality' and `emergence' do not necessitate this type of particle-physics-inspired scale separation; it is the other conditions, such as dynamical separation, that form the essence of emergence.

%\cniels{Dennis: Are the two regimes in our case really adequately described via length scale plus energies? Niels: No: the condition that distinguishes the two regimes has to do with comparing the De Broglie wavelength with the average interparticle distance, and with the time being sufficiently long for the phonons to thermalise/cohere. Mention that this mixes up the clean separation even beyond the 'inverted scale' just pointed out. Do this once we know whether the two superfluid conditions will be mentioned in the FDM section, or whether they will need to be mentioned here.}

%2.2 mixture

\subsubsection{Dynamical separation}

Consider finally dynamical separation. This feature would be violated if the observations require a \emph{dynamical mixture} (or \emph{dynamical dualism}), i.e.\ if the total dynamics depends on non-negligible contributions arising both from the DM nature and from the MG nature of the basic objects of a theory, without the MG contribution being reducible to the DM contribution. One logically possible way in which this might happen is if there are (meta)physical reasons to distinguish two numerically and qualitatively distinct `entities',\footnote{Entities is meant in a broad sense here.} one dark matter and the other a modification of gravity. Call this ontological dualism a \emph{metaphysical mixture}. If these `entities' are equally fundamental, this would resonate with the egalitarian positions. One specific way\footnote{A metaphysical mixture need not be a spatial mixture. Examples are Khoury's two scalar fields (see the conclusion of the prequel paper \cite{martenslehmkuhl1}) \cite{khoury2015} in regions where both fields are non-zero, or the Minkowski metric field and the electric field of a charged point particle in special relativity.} in which a metaphysical mixture might occur is if these two entities are (at least partially) separated in space(time), like a mixture of water and wine. Call this a \emph{spatial} mixture. Given that SFDM introduces only a single field $\Phi$, it seems \emph{prima facie} to be the case that if the DM and MG aspects were to form a dynamical mixture, this would have to be so without them forming a metaphysical let alone a spatial mixture. However, Lagrangian \ref{eqX} only describes the highly idealized case of a pure superfluid at zero temperature. In reality, at finite subcritical temperature, the system is better described phenomenologically by Landau's two-fluid model \cite{berezhiani2016,berezhiani2017}: some sort of mixture of both a superfluid `phase' and a normal `phase'. We would do well to investigate the metaphysics of this model, in order to determine whether it supports a metaphysical or even a spatial mixture.

Landau \cite{landau1941a,landau1941b,landaulifshitz1987}\footnote{Building upon earlier work by Tisza \cite{tisza1938a,tisza1938b,tisza1938c,tisza1940a,tisza1940b}.} developed his two-fluid model (2FM) of superfluidity in response to a tension between several experiments with Helium II.\footnote{See e.g. Dingle \cite{dingle1952} for other such experiments.} One such experiment concerns Helium II flowing through a very narrow capillary without any apparent resistance due to viscosity or friction (as long as the velocity of the flow stays below some critical value) \cite{kapitza1938}. However, if the viscosity of Helium II is measured by the damping of oscillating discs immersed in Helium II at a temperature just below the critical temperature $T_c = 2.17 K$, one obtains a substantive value, albeit one that decreases with temperature below $T_c$ \cite{keesommacwood1938}. Landau's 2FM deals with this ``viscosity paradox'' \cite[p.34]{donnelly2009} \cite[p.13]{schmitt2015} by considering Helium II as a `mixture' (in some sense to be determined) of two parts/components/phases/liquids: a superfluid component, which has no viscosity and carries no entropy and heat; and a normal component, which is viscous and carries entropy and heat. There is negligible interaction between the components. The normal component is responsible for the damping of the discs, which allows one to assign an `effective inertial mass density' to it. An effective mass density for the superfluid component can then be defined by substracting the normal effective mass density from the total mass density of Helium II.\footnote{In the subsequently developed microscopic model underlying the phenomenological 2FM one may derive the effective mass density of the superfluid component independently \cite{schmitt2015}.} The ratio of the densities of each component is a function of temperature only. At $T_c$ the superfluid component vanishes; at $T = 0$ the normal component vanishes. \todo{IGNORE: Does this mean that there are no phonons, and hence it is possible for particle excitations to exist without there being phonons, which would mean that phonons are ontologically dependent on them? Or is it rather the case that at $T=0$ there are neither particle excitations (in a sense) nor phonons?} The capillary experiment is explained by the normal component remaining stationary with respect to the capillary, whilst the superfluid component moves without dissipation---this is called `superflow'. The decrease with temperature of the damping of the oscillating discs is explained by the density of the normal component decreasing with temperature.

What type of mixture do the two components form? Rice \cite{rice1949,rice1950} has argued that, for empirical reasons, these components form a spatial mixture: below $T_c$ the superfluid component has a fibroid structure; just above $T_c$ it starts to appear in small globules. This suggestion has not gained traction in the literature. The received view is that the two components are ``superimposed'' \cite{dingle1952}, or ``interpenetrating'' \cite{feynman1954,salman2013,amigo2017,alonso2018}, i.e.\ not separated nor separable in position space. \todo{IGNORE: hmmm, should I write a footnote about separating the superfluid component by letting it leak out of a capillary. of course, you'll end up with helium atoms, and if they are heated the normal component returns. See Slide 30} Rather than a spatial mixture the components are, in some sense to be specified, two different modes of motion of a single underlying substance. Landau and Lifshitz therefore use the terms `superfluid' and `normal \emph{flow}', rather than `part' or `component' \cite{landaulifshitz1987}. 

Is the 2FM then best interpreted as a mere metaphysical mixture of two ingredients, i.e.\ are there two numerically distinct entities even though they are not spatially separated? One way to answer this is to get a grip on what is meant by `two modes of motion'. Sometimes the received view is (implicitly \cite{rice1949}) portrayed as the two `components' being separated not in position space but in momentum space. One might even be tempted---Berezhiani and Khoury seem to have fallen for this temptation \cite[p.4]{berezhiani2016}---to identify the superfluid component of the Bose-Einstein condensate (i.e.\ `condensate' in the broad sense) with the particles in the zero momentum ground state (i.e.\ the condensate' in the narrow sense) and the normal component with the particles with non-zero momentum (i.e.\ the ``thermal cloud'' \cite{salman2013}). 

An initial response might be that this would be to commit to a spatial mixture after all: the superfluid component is comprised of the particles at rest, with the normal component being comprised of moving particles at other locations than the particles at rest. This would be the case for a classical system, but not for a quantum liquid. The indistinguishable ground state `particles' of a free BEC are maximally smeared out; they are everywhere. That may still seem to leave open the possibility that the thermal particles are not everywhere, suggesting that there are regions with ground state particle mush, but no thermal particles. However, in order for a liquid to form a (superfluid) BEC in the first place, the thermal De-Broglie wavelength needs to be at least comparable with the interparticle distance (determined from the interparticle potential). There is thus no region where the wavefunction of the thermal particles is negligible. The ground state and the thermal cloud are not spatially separated.

\todo{IGNORE: Connect the 2FM with the non-fundamentalist interpretation (i.e. neutral monism).}
A stronger response is that it is impossible in the first place to identify the two ingredients of the 2FM with the ground state and thermal cloud, respectively. For interacting BECs, at low temperatures including $T=0$, only a small fraction of the particles---roughly 10\% for Helium II---is actually in the ground state, due to the zero point kinetic energy \cite[\S 2.6]{annett2004}. Nevertheless, at $T=0$ \emph{all} the particles participate in the superflow \cite[\S 2.7]{annett2004}, which is the defining feature of the superfluid component. According to Landau and Lifshitz, one cannot assign some helium atoms to one component and other helium atoms to the other component \cite{landau1941a,landau1941b,landaulifshitz1987}---regardless of whether you try to split the atoms according to their location or their momentum.\footnote{Alonso \emph{et al.} \cite{alonso2018} claim that even if it would be possible to assign each Helium atom to a component at a specific time, this assignment would change over time, even in dynamical equilibrium. This claim is superseded by Landau and Lifshitz's claim that one cannot assign Helium atoms to a specific component even at a single instant of time.} 

Then, if the 2FM is a metaphysical mixture, it is highly unclear what type of beasts these two ingredients are. Nevertheless, the viscosity experiments strongly suggest that there are two distinct `things' with metaphysically distinct properties. One thing moves through the capillary at rest while the other stays behind; the latter thing does interact with oscillating disks with the former thing now staying behind. The normal component carries entropy and heat; the superfluid component does not. With each component we can associate a different effective inertial mass density. We might want to conclude that there is a metaphysical mixture of two `things' simply in virtue of their distinct properties, even if we are in the dark as to the carriers of these properties (but we will see below that there are reasons against this).

Another argument that suggests this same conclusion is the following. In the literature a distinction is being made between first and second sound; further empirical support for a metaphysical mixture seems to come from the observation of this second type of sound in the 2FM. First sound, i.e.\ phonons, consists of pressure waves. This means that the total density at each point varies over time. However, if there is a metaphysical mixture of two components each with their own effective mass density, these densities can also wave out of phase, in such a way that the total density stays constant. Since the ratio of the densities of each component is a function of temperature, this corresponds to a temperature wave. This second sound has been observed in Helium II.

These two reasons for the existence of a metaphysical mixture can be criticised. Firstly, although such a mixture indeed implies second sound, second sound does not imply a metaphysical mixture. Second sound can be described purely in terms of temperature, without needing to split the constant density into two components that wave out of phase. 

Secondly, how relevant and metaphysically distinct are the differences in entropy, heat and mass between the two components? One can always and trivially make the non-naturalistic metaphysical stipulation that instead of one entity with a value for the entropy S of $s \in \mathbb{R}^+$, there is one entity with two partial entropies, namely 0 and $s$; or even that there are two co-located entities, one with entropy 0 and the other with entropy $s$. But that one can make that trivial stipulation does not mean that we should; Ockham's razor even speaks against it. That being said, we do observe something coming out of the capillary, and it does not carry entropy, even though the superfluid BEC as a whole does carry entropy. 

Let us turn to the effective inertial mass. If this is indeed a physical mass in the relative sense, a weak equivalence principle---associating with each component a gravitational mass numerically identical to its inertial mass---would ensure that the components have gravitational masses that are (in general) distinct. Whether the weak equivalence principle holds in this novel context would of course need to be tested. But it is not obvious that the effective inertial mass is indeed a physical mass in the first place. That we are able to mathematically define a function of temperature does not mean that this represents a ratio of physical mass densities out there in the real world. For one thing, these supposed mass densities are not conserved (but vary with temperature). Moreover, we know that phonons are in fact massless, in the sense of their dispersion relation being gapless. To be fair, we do know that the Helium atoms have rest masses, which are simultaneously inertial and gravitational masses. But, since we have seen that we cannot say \emph{which} Helium atoms are involved in each component of the 2FM, we cannot simply add their gravitational masses to obtain the gravitational mass of each component.\footnote{However, perhaps it is possible to determine how \emph{many} particles are (and thus how much mass is) involved in total per component.} For the same reason the effective inertial mass of the normal component is not and could not be defined by summing over the inertial masses of the constituent atoms; instead it arises from massless phonons (and higher excitations, called rotons) being excited by the oscillating disks and thereby influencing its inertia. It is not directly clear why gravity would interact with a quantity so defined. Dingle refers to this phonon contribution to the effective mass as non-material \cite[\S4.11]{dingle1952}.\footnote{Dingle takes the effective mass to also include the material mass, i.e.~presumably the gravitational mass of the Helium atoms, but this does not seem to follow from the way the effective mass is calculated.} Feynman states in 1954 that the effective mass (of the normal component) ``is not the average value of any quantity that can reasonably be ascribed to an individual excitation. It appears to have meaning only for the entire group of excitations in, or near, thermal equilibrium'' \cite[p.271]{feynman1954}; and more strongly in 1972: it ``is a derived concept and not the density of anything''\cite[p.318]{feynman1972}. In summary, it is far from obvious that one can defend a metaphysical mixture by claiming that there are two distinct components in virtue of them having distinct entropies (in a non-trivial sense) or them having distinct gravitational masses (in a strong physical sense).

Moreover, even if a metaphysical mixture were defensible, it does not seem to be a mixture of one pure DM component and one pure MG component. The normal component---as the name suggests---acts like a conventional viscous fluid \cite[p.29]{annett2004}, i.e.~like the phase of $\Phi$ (or Helium) above $T_c$ which is interpreted as DM. The superfluid component exhibits the novel features, such as superflow, that sparked the interest in superfluids as opposed to normal fluids. One might thus be forgiven for thinking that the normal component is the DM component and the superfluid component the MG component of the mixture. However, the MG behaviour has nothing to do with superflow. And it is the normal component that carries the phonons. After all, it is in virtue of these and higher excitations that the normal but not the superfluid component increases the effective inertia of the oscillating disk, and that the superfluid but not the normal component flows through the capillary without dissipation. The phonon-mediated MOND force is thus associated with the normal component. Can we then interpret the superfluid component (only) as dark matter? 
%The association of an effective mass with each component might seem to suggest that both are DM-like, but we have already questioned whether this is in fact a physical mass. Moreover, to be dark matter would require having gravitational mass, and, as pointed out, the operational definition of the effective mass corresponds to inertial mass only. If we could distinguish which particles where part of which component, we could simply add their gravitational masses to obtain the gravitational mass of each component, but we cannot so distinguish these particles. Perhaps there is nevertheless some way of determining the total non-zero number of particles associated with each component, even if we cannot say which particles. 
On our criterion for dark matter this is equivalent to asking whether we can associate gravitational mass with the superfluid component (only). In the previous paragraph we have seen why one may be skeptical that the effective inertial mass of a component ensures that it has a gravitational mass. If this skepticism is unwarranted, both components would (in general) have a gravitational mass, not just the superfluid component. If the skeptic is right, we can only associate a gravitational mass with the whole system, not with each component separately. Either way, it is not the case that the superfluid component is purely DM-like and the normal component not at all so. 

In summary: we have investigated what kind of ontology is being described by Landau's two-fluid model of superfluidity. A spatial mixture is not very plausible, nor is it obviously a metaphysical mixture---although the phenomenology still very strongly suggests that there must be such a mixture---and definitely not of one purely DM-like component and another purely MG-like component. It is only ``formally'' \cite[p.114]{dingle1952} \cite[p.15]{schmitt2015} or ``artificial[ly]'' \cite[p.272]{feynman1954} that a distinction between two components may be made; ``it is no more than a means of expression convenient for describing the phenomena'' \cite[p.357]{landau1941b} \cite[p.515]{landaulifshitz1987}. If the condition of dynamical separation is to be violated, it must be because of a mere dynamical mixture.

Is there then a dynamical mixture of dark matter and modified gravity in galaxies? That is, does the total galactic dynamics depend on simultaneous, non-negligible contributions arising both from the DM nature and from the MG nature of $\Phi$, without the MG contribution being reducible to the DM contribution? Yes, it does. The MONDian features arise from the phonons coupling \emph{directly} to the baryons (\hyperref[eqTeVeSlike]{Eq.\ref{eqTeVeSlike}}), not from the gravitational mass of $\Phi$. $\Phi$ of course still has a mass, which means that its gravitational contribution needs to be \emph{added}\footnote{The relative importance of each contribution depends on the values of the parameters. In ref.~\cite[p.3-7,11-12]{berezhiani2017} the DM contribution is more important than in ref.~\cite[p.13]{berezhiani2016}. In fact, in the latter paper the DM contribution is considered negligible for radii much smaller than a certain transition radius. However, this does not change the fact that the remaining phonon contribution is not reducible to the gravitational mass of $\Phi$. Moreover, near the transition radius both contributions do become important.} to the phonon contribution. Berezhiani \emph{et al.} refer to this as the hybrid method of calculating the total dynamics \cite[p.4,11-12]{berezhiani2017}. This dynamical mixture violates the criterion of dynamical separation/autonomy (i.e.\ the fundamental description is a description in terms of fundamental concepts (i.e.~DM) only, and, ideally, the emergent description is a description in terms of emergent concepts (i.e.~MG) only).

However, a friend of the DM-fundamentalist approach might push back against this: it would indeed have been pleasant if the emergent description were in terms of emergent concepts only, but the only thing that is crucial is that the fundamental description is a description in terms of fundamental concepts only. Perhaps the hybrid method belongs to the emergent description only, with the fundamental description being able to provide a non-MG story underlying the phonon contribution (even though we have seen that that story would also have to be something other than the gravitational contribution of the dark matter mass). If we take $\mathcal{L}_{\Phi}$ (\hyperref[eqBerezhiani]{Eq.~\ref{eqBerezhiani}}) to be the relevant fundamental description, there is indeed no MG-aspect to be identified whatsoever, as there is no direct coupling to luminous matter which might modify the spacetime that luminous matter would `experience' (i.e.~not the Einstein metric). But this leaves only the option of influencing luminous matter indirectly by contributing via the mass of $\Phi$ to the stress-energy tensor that determines the Einstein metric, as would be expected from dark matter. And it is for exactly this same reason that this can hardly be the complete fundamental Lagrangian, since we do require a direct coupling between phonons (arising from $\Phi$) and luminous matter at the effective level, $\mathcal{L}_{eff, \neg rel, int}$ (\hyperref[eqTeVeSlike]{Eq.~\ref{eqTeVeSlike}}), to obtain the desired results for Galaxy rotation curves. It is difficult to see how an emergent Lagrangian describing the direct interaction between phonons and luminous matter could arise from a fundamental Lagrangian that contains no non-gravitational interaction between $\Phi$ and luminous matter whatsoever. Only the non-interacting effective phonon Lagrangian \ref{eqX} emerges. The creators of SFDM admit that $\mathcal{L}_{eff, \neg rel, int}$ is an ``empirical term'' \cite[p.8]{berezhiani2016} that is added to the phonon description to obtain the desired results for Galaxy rotation curves. They suggest that such a coupling may arise if baryonic matter couples to the vortex sector of the superfluid \cite[p.8]{berezhiani2016}, or it may be a non-perturbative effect \cite[p.5]{berezhiani2017}. Or, it ``\emph{could} be a soft \emph{fundamental} coupling between DM and baryons'' \cite[p.5]{berezhiani2017} (italics added). In any case, some interaction term needs to be included in the fundamental Lagrangian; they explicitly acknowledge that ``a fundamental description of [their] DM superfluid is still lacking'' \cite[p.6]{berezhiani2017}; $\mathcal{L}_{\Phi}$ is incomplete. Thus, arguably, the only reason that $\mathcal{L}_{\Phi}$ (\hyperref[eqBerezhiani]{Eq.~\ref{eqBerezhiani}}) does not tell an MG-story underlying the phonon contribution to the total dynamics is because it currently tells no such story at all. Given that a completion would have to ultimately account for the hybrid method, we may well expect this completion to exhibit MG features.

To sum up, although the lack of a complete theory of SFDM makes it difficult to reach a definite conclusion regarding the satisfaction of the dynamical separation criterion---that is, the fundamental description is a description in terms of fundamental concepts (i.e.~DM) only, and, ideally, the emergent description is a description in terms of emergent concepts (i.e.~MG) only---we have good reasons to doubt that this criterion is satisfied.

This then brings us to a summary of our discussion of the fundamentalist interpretations. For each aspect of $\Phi$, DM and MG, is this aspect fundamental or emergent? \todo{IGNORE: Should I here also summarise the stuff related to their definition of DM as the particle phase?} On our understanding of these aspects they are not both distinct phases of $\Phi$ but two different types of beast. The DM aspect is the massiveness of $\Phi$, the MG aspect is mediated by phonons carried by the normal component of $\Phi$ when $\Phi$ is in the condensed phase below the critical superfluid temperature. 

Within the theory as described by \hyperref[tempting]{Lagrangian~\ref{tempting}} the mass is a fundamental parameter. Since the free $\Phi$ terms ($\mathcal{L}_{\Phi}$, \hyperref[eqBerezhiani]{Eq.\ref{eqBerezhiani}}) are not the reason that this Lagrangian is not the complete Lagrangian, there is good reason to believe that these terms will survive in the unknown complete theory, which would preserve the conclusion that the DM aspect of $\Phi$ is fundamental. 

The MG aspect of $\Phi$ is associated with the superfluid BEC phase (i.e.~the multi-component phase below the critical temperature, as opposed to the single-component phase above that temperature). `Phase' reminds us of thermodynamics, a paradigm example of an emergent theory. If we want to consider all interesting aspects of this phase, we may well have to go to the thermodynamic limit. But the Thermodynamics of SFDM is a different theory from the theory described by \hyperref[fullTzero]{Lagrangian~\ref{fullTzero}}. In our evaluation of $\Phi$ according to the spacetime criteria, no thermodynamics was invoked. \hyperref[fullTzero]{Lagrangian~\ref{fullTzero}} sufficed. Within that single Lagrangian, $\theta$ and $m$ are on a par, suggesting that $\theta$ is as fundamental as $m$. %(Earlier we mentioned that, in the case where a phonon description is dual to an underlying atom or particle description, it may be tempting to consider the phonon description to be the more fundamental description in light of it having more explanatory power. If the particle description were equated with DM, this would constitute MG-fundamentalism. We have now realised however that the antecedent of the previous sentence is incorrect on our understanding of dark matter.) 
This conclusion receives further confirmation from the scale separation and dynamical separation criteria being problematic when applied to SFDM. The similar case of phonons in an atomic crystal suggests that the formal asymmetry criterion may also be violated. The metaphysical asymmetry criterion is however satisfied in some sense.

Thus, the lower half of \hyperref[fundtable]{Table~\ref{fundtable}}, MG-fundamentalism and non-fundamentalism, is ruled out, but both of the views in the top half, (MG+DM)-fundamentalism and DM-fundamentalism, are still contenders for the correct interpretation of $\Phi$ within SFDM. Perhaps a complete theory of SFDM will break the tie, or perhaps the MG aspect of $\Phi$ is simply emergent in some senses but not in others---we will briefly return to this in \hyperref[chartint]{Subsection~\ref{chartint}}.

%Where does that leave the MG aspect? 
%It is a phase. That's a concept within thermodynamics, and hence emergent. But do you really need to use the fact that it is a phase? What we use for the MONDian approach is that there are phonons... and these have nothing to do with the upper-lying description, but feature in the microscopic theory... they are quantum objects. 

\todo{IGNORE: NOTES: Niels, also mention somewhere that it is supposedly not so that the emergent description only applies in galaxies and is inadequate elsewhere. It is simply the case that outside of galaxies the modification of the Einstein metric is 0 (supposedly, but discuss this in the SFDM section, Niels) and hence the effective metric and Einstein metric are the same. So in a sense the phonon description applies everywhere... it's just that it is trivial (i.e. 0) outside of galaxies. That doesn't make it emergence. Phi itself will also be (approximately) zero in places.}

\section{Breakdown interpretations} \label{breakdown}

\todo{IGNORE: Thinking out loud: if one of the elitist fundamentalist interpretations is/were true, would this provide a sufficient difference between the two categories to count as 'work' that blocks these distinctions from breaking down? Or is it the other way round: if the double fundamentalism is true, then the distinctions are in some sense fundamental....?}

%\todo{Yes, spacetime still gives geodesic and in that sense an inertial non inertial distinction (although no gravity inertial distinction). But only for luminous matter, not for Phi. So are the spacetime criteria thus not satisfied for phi test particles? No. Is this a problem? Is this a circularity? Don't know. So yes, spacetime does work by providing the distinction for luminous matter, but not for all matter}

%\todo{Does the blue/green/blircular story refer to the strong inapplicability breakdown approach rather than the weak one?}

We turn to breakdown interpretations, a weak version of which we sympathise with in the context of SFDM. The four interpretations in this group have in common that they take the conceptual distinction between (dark) matter and (modified) spacetime to break down, albeit for different reasons and in different ways. Rynasiewicz has argued that the history from Newtonian physics to GR, via the concept of the aether and the development to field theory, already breaks down the distinction between matter and space(time) as it was classically conceived \cite{rynasiewicz1996}, but he does not distinguish between four different ways in which a breakdown can happen. We will pry them apart, since only the weakest version seems to apply to SFDM---and it may be expected that other theories satisfy some of the other interpretations but again not all of them.

\subsection{Coherence}

%\todo{WRite footnote that some people think that Trinitarianism is incoherent. Cite a textbook on theology that says that. \url{https://plato.stanford.edu/entries/trinity/}}
First consider three strong versions of a breakdown interpretation. On the strongest of these three versions, the DM--MG distinction and by extension the matter--spacetime distinction is considered \emph{incoherent}.\footnote{The consubstantiality thesis/ Trinitarianism (\hyperref[jesus]{Section~\ref{jesus}}) has similarly been accused of being incoherent \cite{septrinity}.} Rynasiewicz seems to hint at such an interpretation when he summarises his argument as hoping ``to have established that the substantivalist-relationist controversy is not necessarily well formulated in every theoretical context'' \cite[p.293]{rynasiewicz1996}, and when he claims that ``[i]nsofar as [the substantivalist-relationist controversy] is intended to have anything to do with physics, ... it is no longer ... \emph{meaningful}'' \cite[p.279, italics added]{rynasiewicz1996}. On this incoherentist strong breakdown interpretation, the SFDM scalar $\Phi$ seeming to be both maximally matter-like and maximally spatiotemporal is taken to reveal that it is in fact incoherent to distinguish these concepts in the first place. This follows, for instance, if one takes the well-known container metaphor seriously. One might phrase this metaphor as ``spacetime is the container in which matter is contained'' or, on a functional understanding: ``to be spacetime is to contain and to be matter is to be contained'' \cite[p.161-167]{sklar1974} \cite[p.306]{rynasiewicz1996}. It is indeed incoherent to say that $\Phi$, playing both the spacetime and matter role, is `contained in itself', since containment is an irreflexive relation. The main argument of the prequel paper would become a reductio ad absurdum. 

Another motivation\footnote{If spacetime and matter are the only two available categories, the container metaphor becomes a special case of the strict dichotomy assumption.} for this version of the strong breakdown interpretation could be the assumption that it is essential to a conceptual spacetime--matter distinction that it is a strict dichotomy: every object is either (an aspect of) spacetime or (an aspect of) matter but never both. It seems to be the case that this is what Rovelli has in mind when he claims, in the context of GR, that the ``very distinction between spacetime and matter is likely to be ill-founded \cite[p.181]{rovelli1997}''. Rynasiewicz similarly considers whether the breakdown of the spacetime--matter distinction in the context of the electromagnetic aether arises from the aether seeming to fall into both categories \cite[e.g.~p.286,290]{rynasiewicz1996}. In any case, if this strict dichotomy were required, $\Phi$ would be a straightforward counterexample.

However, none of the criteria in our two families is necessarily motivated by the container metaphor nor requires these families to constitute a jointly exhaustive and mutually exclusive dichotomy. Nothing in the prequel paper depended on these motivations. 
%The two families of criteria provide an instance of a coherent attempt to define the concepts of matter and spacetime. This is a conceptual distinction between two possible features that an object might have or roles that it might instantiate, with no problem of incoherence or arising from $\Phi$ playing both roles---elements of one family do not in any way contradict elements of the other family.
That being said, it is important to note that the two families of criteria are not totally unrelated. Within the family of spacetime criteria, the (strong) geodesic and chronogeometricity criteria (\hyperref[spacetime-matter]{Section~\ref{spacetime-matter}}; \cite[\S 3.2]{martenslehmkuhl1}) refer to matter, namely in the form of test particles, light rays, rods and clocks. That by itself does not imply any inferential circularity and hence incoherence of the spacetime--matter distinction: one simply starts by identifying the matter fields, and then uses those as input for the references to matter in the criteria for spacetime. However, in the prequel paper only test particles, rods and clocks consisting of luminous matter were considered; in other words, consisting only of pure matter, i.e.\ fields that are only matter \emph{and not also spacetime}. It may thus seem that in order to evaluate the criteria for being spacetime in the context of SFDM, one already needs to know beforehand which matter fields are \emph{not also spacetime}; inferential circularity looms. 

In practice there is however no such inferential problem. In the context of SFDM, one starts by identifying all the matter fields without yet caring whether they are pure or hybrid matter. These are the field $\Phi$ and the luminous-matter fields $\psi^{\alpha}$. One then inspects the Lagrangian to determine how these fields couple to the metric fields in the theory. In the superfluid regime, the $\psi^{\alpha}$ couple to the physical metric $\tilde{g}_{\mu \nu}^{SFDM}$, which is constructed from the Einstein metric and $\Phi$ (\hyperref[crucial]{Eq.\ref{crucial}}). There is no analogous term, $\mathcal{L}_{matter}(\g^{SFDM},\Phi, \Phi_{;\mu | \tilde{g}^{SFDM}})$, for $\Phi$ qua matter field; the free field Lagrangian for $\Phi$ (see eqs. \ref{eqBerezhiani} \& \ref{tempting}) refers (implicitly) to the Einstein metric instead. Thus, $\tilde{g}_{\mu \nu}^{SFDM}$ satisfies the spacetime criteria with respect to the matter fields $\psi^{\alpha}$ and not with respect to the matter field $\Phi$. Moreover, in SFDM no further metric appears that is constructed from the $\psi^{\alpha}$ fields. All this suffices to \emph{simultaneously} make the following two claims: 1) the most spacelike object in the theory (in the superfluid regime) is $\tilde{g}_{\mu \nu}^{SFDM}$ (which, remember, is constructed from $\Phi$), and 2) the $\psi^{\alpha}$ are pure matter. This leaves open two options, neither of which imply an incoherent spacetime--matter distinction. Either (1) is decided to be sufficient to call $\tilde{g}_{\mu \nu}^{SFDM}$ spacetime, making $\Phi$ a hybrid object, and the $\psi^{\alpha}$ the only pure matter fields (with respect to all of which $\tilde{g}_{\mu \nu}^{SFDM}$ satisfies the spacetime criteria). Or, (1) is considered insufficient to call  $\tilde{g}_{\mu \nu}^{SFDM}$ spacetime, from which it follows that there is no spacetime at all in the theory; $\Phi$ and $\psi^{\alpha}$ are all pure matter fields. Hence, in the latter scenario the spacetime category would not be ill-defined, but just inapplicable to SFDM. This would make SFDM amenable to the next interpretation to be discussed, not the incoherence interpretation. 

In the prequel paper the first option was chosen. This may seem ad hoc. Isn't the point of spacetime that it is `experienced' universally by all matter? That intuition stems from times when all theorised matter was pure matter. It is silent on the question of how to extend the concept of the universality of spacetime to theories with (tentatively) hybrid matter. If there is an option that ensures (coherence and) applicability and an option that does not, isn't the charitable thing to do to go with the former? (That is, assuming application of the spacetime category is at all useful---this will be questioned below, as uselessness is much weaker a charge than incoherence or even inapplicability.)

%It is not the case that $\tilde{g}_{\mu \nu}^{SFDM}$ satisfied the criteria for being spacetime with respect to test particles, rods and clocks made out of $\Phi$. The free field Lagrangian for $\Phi$ (see eqs. \ref{eqBerezhiani} \& \ref{tempting}) refers (implicitly) to the Einstein metric, not to $\tilde{g}_{\mu \nu}^{SFDM}$.

\todo{IGNORE: Hmmm, the matter criteria also refer to spacetime, via the stress-energy momentum tensor (see Dennis' stress energy momentum paper ). ik: well, it refers to a metric field. It doesn't need to identify it with spacetime per se. Maybe discuss this in another paper. Dennis says that it is all very subtle; for instance, the spacetime criteria also don't assume that test particles are matter... its just that the Weatherall stuff shows that test particles follow geodesics, which we may then interpret as sufficient reason for dubbing them matter.
Two miracles Paper, fn37, both sides of EFE require metric to be defined. }

\subsection{Applicability}

The second-strongest breakdown interpretation grants the coherence of the (dark) matter and (modified) spacetime categories, but claims that they are inapplicable to the theory or field under consideration---for reasons beyond the previous uncharitable claim of the inapplicability of the spacetime category. On this interpretation, there is nothing inherently wrong with these concepts, but it would be a category mistake to apply them to $\Phi$. It is like finding out that one should not use Aristotelian elements---earth, air, fire, water, aether---but Newtonian massive point particles, or, different still, quantum fields. Perhaps this inapplicability of the traditional concepts of matter and spacetime is the core of Rynasiewicz's interpretation of the history from Newtonian physics to GR:
\begin{quote}
	[O]ne could just as well say that [the aether] is neither [space nor matter], but instead belongs to a category of its own. \cite[p.290]{rynasiewicz1996}
\end{quote}
\begin{quote}
	In the course of its development, physical theory simply lost touch with the categories necessary for the original formulation of the [substantivalist-relationist controversy, i.e.~space(time) and matter]. \cite[p.279]{rynasiewicz1996}
\end{quote} 
\begin{quote}
	Present-day physicists do \emph{not} employ a language that conforms with the original contrast [between matter and space(time)] \cite[p.301, italics in original]{rynasiewicz1996}\footnote{See also \cite[p.80]{rovelli2010}.}
\end{quote}
This raises an obvious question: what, if any, are the correct categories? Rynasiewicz and Earman do raise this question, affirming that there must be other such categories, but they do so only in the last paragraph of their paper and book, respectively, leaving it open what exactly the alternative option is supposed to be:
\begin{quote}
Some have suggested, as has Earman [see e.g.~\cite[p.208]{earman1989}], that what is needed now is a
tertium quid to the traditional positions, something that will require an act of scientific creativity. I submit that this act, or rather, series of acts, has already occurred, but, for a multitude of reasons, we have simply failed to recognize it. \cite[p.306]{rynasiewicz1996}
\end{quote}
It is important to point out though that, while the old categories may well be \emph{replaced} by new categories, there is no guarantee that nor any urgent reason why there would be new categories: the inapplicability of the old categories may simply indicate that they are \emph{eliminated} rather than replaced. %This latter view resembles that of the mature Einstein, according to whom the metric field and paradigmatic `matter' fields such as the electromagnetic fields are all just fields, with any further distinction, such as that between matter fields and other fields, being unimportant, or even misleading \cite{Lehmkuhl:forthOUP}.

Does this inapplicability interpretation apply to SFDM? Are the concepts of matter and spacetime simply the wrong conceptual tools to understand this theory? If so, what glasses, if any, should we use? What is this tertium quid, tentatively labeled ``dark stuff''? Should we modus tollens the modus ponens of the prequel paper---`if the families of criteria for spacetime and matter are applicable, then $\Phi$ satisfies both'? We do not think that the inapplicability thesis is a viable interpretation of SFDM. For one thing, the category of (`pure') matter still applies---to the luminous matter fields, \hyperref[crucial]{Lagrangian~\ref{crucial}}, which are only matter and not also an aspect of spacetime. Similarly, the category of (`pure') spacetime is still expected to apply in the final SFDM theory to the Einstein metric in the regime $T \gg T_c$ where we expect any modifications to be negligible (see the end of \hyperref[FDMintro]{Section~\ref{FDMintro}}). Moreover, it is not the case that the proposed new category, ``dark stuff'', is unrelated to the traditional categories. It is fairly directly related, by being the conjunction of both traditional categories. It is a hybrid category, rather than a novel, irreducible category. Hence, this case is not at all analogous to, say, the move from using only Aristotelian elements to using only strings. The inapplicability breakdown interpretation does not apply to $\Phi$ within SFDM, or at least not beyond the extent one may consider it to hold already in the case of plain GR. The inapplicability interpretation would be more appropriate for objects that belong to neither category than for objects that belong to both categories.

\subsection{Objectivity}

The final strong breakdown interpretation does grant that there is a coherent conceptual distinction between (dark) matter and (a modification of) spacetime that applies to the field under consideration, but it is a conventional, rather than an objective fact, whether that field instantiates one or the other aspect. This interpretation is of particular interest in light of the claimed duality between f(R) gravity and particular Brans-Dicke-Theories and the further claimed equivalence between the Einstein and Jordan frames/representations of Brans-Dicke Theories \cite{duerrms}.\footnote{Similar questions of conventionalism seem to arise in string theory in the context of the claimed equivalence between the string frame and the Einstein frame \cite{alvarez2002} (which are related by a conformal transformation involving the dilaton). We would like to thank James Read for suggesting this interesting comparison.} Rynasiewicz \cite{rynasiewicz1996} hints that the distinction between space(time) and matter is conventional in a much larger context---and without even requiring the existence of a duality or equivalent representations---referring to it as a ``verbal [dispute] occasioned by arbitrary preference''\footnote{This is how North interprets Rynasiewicz \cite{north2018}.} (p.279), a ``\emph{fa\c{c}on de parler}'' (p.287; italics in original), a mere ``mode of expression'' (p.290), a matter of ``taste'' (p.290), ``a matter of whim'' (p.299), not a matter of fact (p.301).\footnote{Vassallo refers to the distinction as ``a mere choice of vocabulary'', since it is ``naive and perfectible'' \cite[p.354]{vassallo2016}, and Rovelli states that it is merely a matter of ``semantics'', ``only a matter of choice of words, and thus, ultimately, personal taste'' \cite[p.77]{rovelli2010}, but it is not clear whether they mean that it is conventional, or that it is dispensable (see the weak breakdown approach).} However, it is an objective fact in which phase the field $\Phi$ in SFDM is, and hence an objective fact when it instantiates which role. The conventionalist strong breakdown approach is not appropriate for $\Phi$ as it features within SFDM, or at least not more so than it may already be in plain GR.

%ryno: weak breakdown: ``The fundamental conclu- sion I am urging-that the substantival-relational debate is no longer a fruitful one for contemporary natural philosophy-re- mains unaffected.''

\subsection{Utility}

This leads us to a weak version of the breakdown interpretation. The matter and spacetime categories \emph{may} coherently and objectively be applied to objects in physical theories, such as SFDM. Up to this point this interpretation agrees with the fundamentalist interpretations and the consubstantiality interpretation. The difference is that whilst on those interpretations this distinction is considered to be useful (and hence must be retained), on the weak breakdown interpretation it is considered to be dispensable. Indeed, there does not seem to be a clear benefit provided by these categories, at least within SFDM. They are no longer useful, or, as Rynasiewicz puts it, no longer ``fruitful'' \cite[p.291]{rynasiewicz1996}. It is then only a small further step to abandon such surplus vocabulary, on the grounds of conceptual parsimony.

What could be a reason to retain these concepts? We have already seen two motivations---the container metaphor and viewing the spacetime--matter distinction as the most basic metaphysical dichotomy such that everything falls into exactly one of these categories---which, if they ever were a motivation for introducing the spacetime--matter distinction, definitely can not serve as reasons anymore in light of SFDM. One might insist that we still need this distinction to make sense of the (modified) Newtonian gravity and special relativistic regimes of theories such as SFDM and GR. These concepts may well be useful in those limits, just as the concepts of chairs, tigers and temperature are, but that does not make these concepts fundamental, i.e.~relevant for the fundamental description of reality, just as chairs, tables and temperature are not. 
%\todo[color=white]{Dennis, have you perhaps read Ladyman and Ross's 'every thing must go'? If so, is that of relevance for our discussion?}

Mentioning Newtonian Gravity just now does remind us of the `work' that the concept(s) of spacetime (and matter) did in that theory: space(time) provided a distinction between inertial and non-inertial motion (of all matter). Hence, when the (strong) geodesic criterion (\S \ref{spacetime-matter}) is included in the set of criteria that allow interpreting something as spacetime, it may seem that the weak breakdown interpretation becomes analytically false: if an object is spacetime with respect to test particles of matter in the sense that the trajectories of all those test particles follow the geodesics of that object (and non-test particles do not) then that concept of spacetime so instantiated automatically does `work' in that it distinguishes between inertial and non-inertial motion of all matter, and hence the concept deserves to be retained. But this is only correct in a restricted sense. The qualifiers `all' are crucial. As mentioned earlier in this section, the prequel paper concluded that $\tilde{g}^{SFDM}$ satisfies the criteria for being spacetime with respect to test particles (and rods and clocks) made out of luminous matter, which is pure matter, i.e.\ fully matter but not at all spacetime. But what about test particles made out of $\Phi$? They couple to (only) the Einstein metric rather than to $\tilde{g}^{SFDM}$ and hence the geodesic criterion is not satisfied for the combination of $\tilde{g}^{SFDM}$ and $\Phi$-test-matter. And it's exactly the conceptually novel $\Phi$ field that we are trying to interpret. And it is not the case that $\Phi$, qua matter, experiences what we had identified as spacetime ($\tilde{g}^{SFDM}$) nor is it true that $\Phi$, qua aspect of spacetime (via $\tilde{g}^{SFDM}$), provides a distinction between inertial and non-inertial motion for \emph{all} matter. Thus, this specific type of `work', that we are used to being done by the concept(s) of spacetime (and matter), does not extend fully to hybrid theories such as SFDM.

%(In fact, one may insist that this invalidates the conclusion of the prequel paper, and strictly speaking it does. If one considers it an essential feature of spacetime that it is universal in the sense that all matter experiences the same single spacetime, as encapsulated in the geodesic criterion (ranging properly over \emph{all} test matter), then no single object and hence no object at all in SFDM is strictly speaking spacetime (cf.\ fn.\ref{multipleST}). This is of course grist to the mill of the breakdown approaches, rather than an objection to it.) \todo{Incoherence interpretation.}

%NOTE TO SELF: point out that this may seem to invalidate the main conclusion of our prequel paper, namely that g-tilde is spacetime. Sure, this is true to some extent. But it then follows that nothing is spacetime in that theory (also not the Einstein metric), so this only strengthens the breakdown interpretations. In fact, don't make this a footnote, but mention it as an additional paragraph as an argument in favour of the breakdown interpretations. Also write footnotes in the prequel paper referring ahead to this caveat of not satisfying the spacetime criteria for all (test) matter.... (in order words, say that there was a caveat when I used 'test matter' and 'rods and clocks' etc).

Alternatively, in response to the question of what reason one may have to retain the concepts of spacetime and matter, one might consider the idea that (modified) spacetime and (dark) matter have different explanatory roles to play.\footnote{Depending on one's favourite model of explanation, the previous suggestion---providing a distinction between inertial and non-inertial motion---may also be an instance of this `alternative' suggestion of considering the explanatory roles of spacetime (and matter).} One example would be the claim that spacetime and matter stand in an asymmetric explanatory relation to each other. For instance, consider theories with a dynamical metric and other fields such that if we create test particles, light rays, rods and clocks out of these other fields, the metric satisfies the (strong) geodesic criterion (i.e.\ if test particles and/or light rays were around they would, for any choice of initial conditions, follow the timelike and null geodesics, respectively, of this metric) and the chronogeometricity criterion (i.e.\ local validity of special relativity) with respect to those other fields (see \cite{martenslehmkuhl1}). We may then want to call the metric `spacetime' and those other fields `matter'. On the so-called `dynamical approach' to this metric, it is \emph{because} of properties of those matter fields and the way in which they are coupled to the metric that the metric field obtains its chronogeometric meaning (and presumably also its inertial meaning, and thereby thus full spatiotemporality) \cite{brown2005}. Properties of matter fields (and the way in which they are coupled to the metric) \emph{explain} why matter---rods, clocks, test particles---surveys the metric field (which therefore justifies calling the metric `spacetime'). On the opposing approach, the `geometrical approach', the arrow of explanation is reversed: the metric is primitively spatiotemporal, in that it \emph{constrains} the properties of the other fields such that those survey the metric field (in the same sense as before, i.e.\ the (strong) geodesic and chronogeometricity criteria). The metric field thus \emph{explains} those properties and thereby justifies referring to those fields as `matter'. 

Does this move apply to SFDM? We contend that it does not, regardless of one's preference for the direction of the arrow of explanation. Consider first whether $\Phi$ might be both spacetime and matter (as indeed suggested by our two families of criteria which were silent on any (further) explanatory asymmetry). In other words, could it be that $\Phi$ is matter and $g_{ab}^{SFDM}(\Phi)$ spacetime in the specific sense of the previous paragraph? This would be guaranteed by a $\Phi$ term in the SFDM Lagrangians analogous to the last terms in Eqs. \ref{fullTzero} and \ref{tempting}, i.e.\ $\mathcal{L}_{matter}(\g^{SFDM},\Phi, \Phi_{;\mu | \tilde{g}^{SFDM}})$. There is no such term. There may be an arrow of explanation between $g_{ab}^{SFDM}(\Phi)$ and luminous matter, but there is no parallel arrow between $g_{ab}^{SFDM}(\Phi)$ and dark matter ($\Phi$). Could we then perhaps view $\Phi$ as being only spacetime but not matter, in the explanatory sense of the previous paragraph? Not only would this be to ignore that $\Phi$ scored maximally on our family of matter criteria, but also, once again, that as far as we know the MG-aspects only appear in the superfluid regime, and we do not yet have a complete and fully satisfactory fundamental SFDM Lagrangian available to determine whether $\Phi$ is in any sense a modification of spacetime at the fundamental level. This specific attempt to justify a (modified) spacetime vs.\ (dark) matter distintion via differing explanatory roles is of no avail. %We will briefly return to one further suggestion along these lines at the end of this section.

Until now we have considered theory-internal senses of the `work' that the categories (dark) matter and (modified) spacetime might be doing. Perhaps their usefulness arises at a more pragmatic, theory-external level: a heuristic role in theory-development. We will discuss this issue---the usefulness of the conceptual distinction between (dark) matter and (modified) spacetime at the level of the space(s) of theories---in \hyperref[upshot]{Subsection~\ref{upshot}}. However, this move from a theory-internal to a theory-external usefulness of the distinction is to declare it dispensable within SFDM considered in itself. As it stands we thus sympathise, within the group of breakdown interpretations, with the weak breakdown interpretation of the DM/MG distinction in the context of SFDM. Nevertheless, one may argue that one specific feeling of unease remains on this interpretation. Historically, theories that have been labeled as modifications of gravity/spacetime have done very well at explaining galactic data but not so well at explaining or even accounting for data at larger scales, and vice versa for theories that have been labeled as dark matter. This may give the impression that one necessarily needs something spatiotemporal to explain the galactic data and something (dark) matter-like to explain data at larger scales. Could it be these distinct explanatory roles that could be used to re-establish a fundamental conceptual distinction between (modified) spacetime and (dark) matter? This is the central question of our next project. On the other hand, perhaps the lesson to be learnt from SFDM is that it is a crucial counterexample to this historical trend, highlighting its contingency.

\section{Cartography} \label{cartography}

%\todo{Use 'chart' instead of roadmap? ``Cartography of the space of theories'' as paper title, and ``cartography'' as title for this section? Atlas? If we change the paper title, change the reference to this paper in paper 1. Or this section: ``What's the matter with spacetime? Cartography of the space of theories}

\todo{IGNORE: Patrick's new JBD paper (section 4.2) seems to suggest that if something is not fundamental, then we can ignore it. So if something is emergently matter and fundamentally spacetime, then it is spacetime simpliciter, and vice versa. Mention this. This would mean that DM-fundamentalism means it really is a DM theory, and hence it would be an argument against the conclusion of the first paper. If only fundamental roles count, then this is just another DM theory (if we ignore that the MG aspect was not emergent in all of the senses). Would this be an extra position? Would this make DM fundamentalism incompatible with both the Jesus and the weak breakdown approaches? Maybe discuss this in the new section 7 where I compare the 3 groups of interpretations. (btw, start that section by saying "the relation between these three groups of interpretations is very subtle. Note, by the way, that Jesus---whether it adds the "and nothing further" clause or not---is incompatible with the strong breakdown approach that calls the categories incoherent. Whether it is compatible with the conventionalist strong and with the weak breakdown approaches does depend on that clause.)}

\begin{table}
	\small
	\definecolor{colorfortablecell}{rgb}{0.77, 0.76, 0.82}
	\hspace{-1cm}
	{\renewcommand{\arraystretch}{1.2}%
	\begin{tabular}{| l | l | c | l | }
	\hline
	{\bf Group} & \multicolumn{2}{l|}{{\bf (Sub)Name}} & {\bf Description}\\
	\hline\hline	
	\multirow{10}{*}{\quad \rotatebox[origin=c]{90}{\parbox[c]{5cm}{\centering  Theological analogies}}} & \multicolumn{2}{l|}{Consubstantiality} & The field $\Phi$ is both fully (dark) matter and fully (a\\ %\begin{sideways}Jesus\end{sideways}
		& \multicolumn{1}{l}{ } & & modification of) spacetime---both roles are on a par.\\
		& \multicolumn{1}{l}{ } & &  This leaves the conceptual categories of (dark) matter \\
		& \multicolumn{1}{l}{ } & &   and (modified) spacetime and their distinction \\
		& \multicolumn{1}{l}{ } & &   unaffected (i.e.\ coherent, applicable, objective and useful).\\ \cline{2-4}
		& \multicolumn{2}{l|}{Demi} & The field $\Phi$ is partially (dark) matter and partially (a\\ %\begin{sideways}Jesus\end{sideways}
		& \multicolumn{1}{l}{ } & & modification of) spacetime. This leaves the conceptual \\
		& \multicolumn{2}{c|}{\textcolor{red}{\bf X}} & categories of (dark) matter and (modified) spacetime \\
		& \multicolumn{1}{l}{ } & &    and their distinction unaffected (i.e.\ coherent, applicable, \\
		& \multicolumn{1}{l}{ } & & objective and useful).\\
	\hline
	\multirow{9}{*}{\quad \rotatebox[origin=c]{90}{\parbox[c]{3cm}{\centering  Fundamentalism}}} & \multicolumn{2}{l|}{(MG+DM)-fundamentalism}  & The field $\Phi$ is fundamentally both (dark) matter\\ 
		& \multicolumn{2}{l|}{(or: Dualism)} & and (a modification of) spacetime. \\ \cline{2-4}
	 
	 	& \multicolumn{2}{l|}{DM-fundamentalism} & The field $\Phi$ is fundamentally (dark) matter, but only\\ 
	 	& \multicolumn{2}{l|}{} & emergently (a modification of) spacetime. \\ \cline{2-4}
		& \multicolumn{2}{l|}{MG-fundamentalism} & The field $\Phi$ is fundamentally (a modification of) space-\\ 
		& \multicolumn{2}{c|}{\textcolor{red}{\bf X}} & time, but only emergently (dark) matter. \\ \cline{2-4}
		& \multicolumn{2}{l|}{Non-fundamentalism} & The field $\Phi$ is merely emergently (dark) matter as well\\ 
		& \multicolumn{2}{l|}{(or: Neutral monism)} & as (a modification of) spacetime; fundamentally it is\\
		& \multicolumn{2}{c|}{\textcolor{red}{\bf X}} & something else altogether. \\ \cline{2-4} 
	\hline
	\multirow{13}{*}{\quad \rotatebox[origin=c]{90}{\parbox[c]{1.5cm}{\centering Breakdown}}} & \multirow{9}{*}{Strong} & \multicolumn{1}{l|}{Incoherence} & The field $\Phi$ reveals that there is no coherent conceptual \\ 
	 & & \textcolor{red}{\bf X} & distinction between (dark) matter and (modified) spacetime. \\ \cline{3-4}
	 & &\multicolumn{1}{l|}{Inapplicability} & The field $\Phi$ reveals that the (dark) matter and (modified)\\ 
	 & & \multicolumn{1}{c|}{\multirow{2}{*}{\textcolor{red}{\bf X}}} & spacetime categories, despite being coherent, are the \\
 	 & &  & wrong conceptual categories for SFDM.\\ \cline{3-4}
	 & & \multicolumn{1}{l|}{Conventionalism} & There is a coherent conceptual distinction beween (dark)\\
	 & & \multicolumn{1}{c|}{\multirow{2}{*}{\textcolor{red}{\bf X}}} & matter and (modified) spacetime, but it is conventional\\
	 & & \multicolumn{1}{l|}{ } &  (i.e. not objective) which role applies to the field $\Phi$.\\ \cline{2-4}
	 & \multicolumn{2}{l|}{\multirow{4}{*}{Weak}} & There is a coherent, objective conceptual distinction\\ 	 
	 & \multicolumn{1}{l}{ } & & between (dark) matter and (modified) spacetime that \\
	 &\multicolumn{1}{l}{ }  & &  applies to the field $\Phi$, but it is dispensable/useless. It is \\
	 &\multicolumn{1}{l}{ }  & & then dispensed with, on grounds of conceptual parsimony.\\
	\hline
	\end{tabular}}
	\caption{Chart of possible interpretations of (theories with) fields that do not fall into exactly one of two putative categories, i.e.\ (dark) matter and (modified) spacetime. A red \textcolor{red}{\bf X} indicates that the interpretation is not a viable interpretation of the scalar field $\Phi$ in SFDM.}
	\label{chart}
\end{table}

In the prequel paper it was argued that the novel field $\Phi$ postulated by Berezhiani and Khoury's `superfluid dark matter theory' not only maximally fits the putative conceptual category of (dark) matter, but it also maximally fits---at least in the superfluid regime---the putative conceptual category of (a modification of) spacetime. This paper has proposed a chart of nine---or ten, if the demi interpretation is included for the sake of completeness---\emph{prima facie} possible interpretations of such Janus-faced objects and their theories, structured into three groups (\hyperref[chart]{Table \ref{chart}}). A first evaluation of those interpretations for the specific case of the scalar field $\Phi$ featuring in SFDM has left four of those options on the table as viable interpretations of $\Phi$: the consubstantiality interpretation, the two fundamentalist interpretations that take at least the DM aspect to be fundamental, and the weak breakdown interpretation. 

%\textcolor{red}{OVERVIEW OF THE NEXT FEW SUBSECTIONS}

%The consubstantiality interpretation holds that the senses in which $\Phi$ is (dark) matter and (modified) spacetime are on a par. There is nothing more to be said. If there is anything objectionable about this interpretation, it would be that it is incomplete. The other groups of interpretations are offered in an attempt to provide a more complete interpretation.

%The fundamentalist interpretations consider for each of the two roles played by $\Phi$ whether these are fundamental or merely emergent aspects of $\Phi$. The matter aspect of $\Phi$ is straightforwardly fundamental, and this would plausibly remain the case if a complete SFDM theory is found. The modified gravity aspect is less straightforward, as it is emergent in some senses but not in others. A definite conclusion is hindered by the lack of a complete theory.

\subsection{A chart of interpretations} \label{chartint}

If such cartography of the space(s) of theories is to present us with a description and ultimately an understanding of the space(s) of theories, we need not a mere list of interpretations, but a full chart that includes how they are `located' relative to each other. While all these interpretations are mutually inconsistent, some pairs are closer in spirit to each other than other pairs. This subsection explores the subtle relations between the various interpretations. 

One might think that the consubstantiality interpretation is nothing more than a repetition of the result of the prequel paper, and that it is therefore the basis of all the other eight interpretations (besides the demi interpretation). The latter statement does not follow, since the incoherence and inapplicability versions of the strong breakdown interpretation disagree with the results of that paper. Moreover, the consubstantiality interpretation has more to it, making it inconsistent with all the other interpretations. For one thing, it insists that the DM/MG distinction is objective and useful, contra the whole class of breakdown interpretations. How does it relate to the fundamentalist interpretations? On the consubstantiality interpretation, the senses in which $\Phi$ instantiates the DM and MG roles are on a par. It is thus in some sense closer to the spirit of the egalitarian interpretations than the elitist interpretations. This could easily be mistaken as agnosticism about the egalitarian interpretations---as long as one of those is viable, the consubstantiality interpretation is viable. Similarly, an agnostic version of the demi interpretation is consistent with any of the four fundamentalist interpretations. In order for all ten interpretations to be mutually exclusive, we focus on versions of the consubstantiality and demi interpretations that embrace quietism about the fundamental vs.\ emergent nature of the DM and MG aspects of $\Phi$. There is then simply no fact of the matter about the fundamentality of the DM and MG roles, just as one might hold that it is not a meaningful question to ask whether Jesus is more, equally or less fundamentally divine than he is human. 

The decision between the consubstantiality, DM-fundamentalist and dualist interpretations of SFDM thus depends on whether there is a matter of fact about the fundamentality of each of the two roles that $\Phi$ instantiates. That the latter two interpretations have not yet been ruled out may be seen as ruling out the consubstantiality interpretation. Especiallly their agreement that the DM role is a fundamental aspect of $\Phi$ may seem funest for the consubstantiality interpretation. However, it is not so much the case that both the DM-fundamentalist and dualist interpretations are unproblematic, but rather that the truth, if any, seems to lie in between. The MG aspect of $\Phi$ fits some of the aspects of ``emergence'', but not others. Perhaps the complete theory of SFDM will solve this. If not, it may be that a different understanding of the concept of emergence is needed, or that a new hybrid fundamentalist view is required, or maybe we should take this problem to vindicate the consubstantiality interpretation, although that seems in tension with the unproblematic claim that the DM aspect is fundamental.

All this being said, there is one sense in which the consubstantiality and fundamentalist interpretations are closely related to each other when contrasted with the breakdown interpretations. The former but not the latter consider the DM/MG distinction to be coherent, applicable, objective and useful. Within the group of breakdown interpretations, the one that is closest to the other five is nevertheless the only one that is still viable: the weak breakdown interpretation. In contrast to the other breakdown interpretations, this interpretation does allow one to use the DM/MG distinction (non-conventionally). On this interpretation it is \emph{allowed} to ponder the question of whether each role is instantiated fundamentally or emergently, but this would not be more than a fun academic exercise. The issue is moot, it is pointless, if the distinction is not useful. 

The decision between the weak breakdown interpretation on the one hand, and the other three remaining viable approaches on the other hand, depends on the usefulness of the DM/MG distinction. In the discussion of the weak breakdown approach, several proposals for the use of the DM/MG and spacetime--matter distinctions within SFDM have been argued against. The onus is thus on the other interpretations to provide a proposal for the usefulness of these distinctions within SFDM. We have hinted that distinct explanatory powers, for instance with respect to galactic data vs data from clusters and cosmology, may provide such a role. Without such a concrete theory-internal proposal however, the contest between the weak breakdown interpretation on the one hand, and the other three remaining viable approaches on the other hand, is perhaps most interesting at the level of the space(s) of theory, as will be discussed in the next subsection.

Note finally the connection between the inapplicability version of the breakdown interpretations and neutral monism. Both agree that the categories are inapplicable at the fundamental level. Their difference though is that this is where it stops for neutral monism, which still acknowledges their applicability at the emergent level, whereas the inapplicability applies universally according to the inapplicability breakdown interpretation. It is this difference that makes the former compatible with the conclusion of the prequel paper, but not the latter.

\subsection{The upshots} \label{upshot}

What are the upshots of all of this? More precisely, what are the upshots of this general chart of interpretations and what is the relevance of applying it to SFDM specifically?

Perhaps the most important upshot is the realisation that there is a single space of theories, rather than a space of spacetime theories and a space of matter theories.\footnote{Even if, for independent reasons, there would be multiple spaces of theories after all, it is still the case that none of those is a pure space of spacetime theories or a pure space of matter theories, nor do they reduce to one of these pure spaces.} This single space is not reducible to (just) those two separate spaces. This is something that could have been realised directly from the two families of criteria for being matter and being spacetime, respectively, but which had not been sufficiently noticed due to the widespread focus on fields that are (thought to be) purely matter or purely spacetime. (Which raises the question: which other conceptual mistakes have we made by focusing only on our favourite theories?) Interpreting SFDM has helped highlight this realisation, which was already noted in \hyperref[breakdown]{Section~\ref{breakdown}}: the distinction between (dark) matter and (modified) spacetime is not a strict dichotomy, i.e.~the distinction does not necessitate that all objects fall into exactly one of these categories. Why did we introduce two independent families of criteria for spacetime and matter respectively? If spacetime and matter are supposed to form a strict dichotomy, would we then not expect a single criterion such that if the criterion is satisfied by an object, then that object is a spacetime, and otherwise it is matter, or vice versa?\footnote{The case for a strict dichotomy gets even worse if we move away from necessary and sufficient criteria for interpreting something as matter or spacetime, but instead consider these categories to be cluster concepts \cite{bakerstfunc}.} There is no reason to expect that an object satisfying any of the spacetime criteria does not (or does) satisfy any of the matter criteria, and vice versa. There is no conceptual link between the two families of criteria, beyond the reference of the spacetime criteria to matter (e.g.~test particles, rods and clocks). In the space of theories, theories with objects that fall under exactly one family of criteria may well be the exception; the \emph{prima facie} expectation should be to also encounter objects that resemble both matter and spacetime, or neither. History has been such that two early theories, Newtonian physics and special relativity, were exceptions in this regard, rather than the norm. We should not be mislead into believing in a strict spacetime--matter dichotomy merely because of our familiarity with these early theories.

In other words, the urge to project pure spacetime and pure matter categories onto theories, such as for instance SFDM, and insisting that everything falls into exactly one of these categories, is an artefact of our familiarity with Newtonian theories and Special relativity. Had SFDM been the only theory we had known, insisting on applying these pure categories would have seemed as strange as someone insisting that we use Goodman's bleen and grue \cite{goodman1955} to describe trees and the sky rather than blue and green---there is nothing incoherent about using bleen and grue, but they just are not particularly useful concepts. If the first few theories or observations we came across had been such that all red things were square or triangular and all blue things circular, we might have developed terms such as `POLYGON' and `BLUE' for these respective groups of objects. If there were no observations of further objects not falling into these categories, we might have surmised that the most basic conceptual distinction is the strict POLYGON-BLUE dichotomy. However, once we would make further observations and/or further explore the space of theories, we might realise that there is no reason why red things could not be circular, and blue things squared or triangular, and that the POLYGON-BLUE pattern in our first theory/data was an exception rather than the norm. We would immediately realise, if we hadn't already (and of course we could have, simply by analysing the concepts of polygon and blue), that this distinction, however coherent, is not particularly useful for fundamental metaphysics---it does not track anything deep about the space of theories. In particular, we would realise that there is a single space of theories with theories/objects that are/contain blue squares, blue circles, red squares, red circles, partially blue and partially red triangles, etc\footnote{One might retort that this leaves open the possibility that there is a different distinction that is the most basic conceptual distinction, for instance one based on size vs.\ hue. However, it is hard to see why that wouldn't face the same problems as the POLYGON-BLUE dichotomy.}---rather than all theories falling either into a space of POLYGON theories or a space of BLUE theories.
%\todo{Connect this stuff to inapplicability breakdown interpretation, and maybe to other breakdown interpretations? Hmmm, it indeed sounds a lot like the inapplicability approach. Maybe rephrase it as "the urge to project a strict dicthotomy is an artefact of our familiarity with....", after all isn't my main point that we need to move to a space of theories rather than two spaces? Niels, can't that point already be made even on the Jesus and fundamentalist interpretations, simly in virtue of the conclusio of paper 1?}

These two companion papers contribute to the `cartography research programme'---a generalisation of the research programme previously advocated by one of us for the space of spacetime theories \cite{lehmkuhl2017}---which aims to help us navigate the single space of theories. Analysing individual theories and the similarities and differences of neighbouring theories helps us to better understand the space of theories as a whole, which in turn helps us better understand further individual theories and pairs of theories, and also to re-evaluate familiar theories, such as GR. The cartography research programme advocates a dynamic back-and-forth between both levels of analysis.
%Within the region of pure spacetime theories, there are three schools of cartographers. The first school has focused on
%the focus has been on......\textcolor{red}{ADD THE TWO OR THREE SCHOOLS OF THOUGHT IN THE LEHMKUHL INTRODUCTION TO THE BOOK.} 
Here we developed a new dimension to the cartography of the space of theories, distinct from the focus of previous cartographers \cite[\S 2]{lehmkuhl2017}, namely interpretation of theories in terms of a DM/MG distinction, as well as the broader spacetime--matter distinction. The resulting chart of interpretations may help understanding further theories (better), whether they are new theories or plain old GR. Moreover, as the chart is not claimed to be exhaustive, analysis of other theories will likely enrich the chart. 

Note that the struggle between the consubstantiality interpretation and the two remaining fundamentalist interpretations on the one hand, and the weak breakdown interpretation on the other hand, may be most interesting at the level of this space of theories. Within the bible, Jesus is the unique human/god entity. It is in virtue of this rareness that it makes sense to keep using the categories of (pure) human and (pure) god. If Janus-faced objects such as $\Phi$ in SFDM turn out to be equally rare in the space of theories, then, even though there is in principle still a single irreducible space of theories, in practice it can be pulled apart into a space of spacetime theories and a space of matter theories.\footnote{Or, if, say, DM-fundamentalism is applicable to the majority of theories, one may be tempted to conclude that, in practice, only the (dark) matter category is conceptually fundamental and that the single space of theories is, basically, a space of matter theories. Note, however, that it is not obvious that a difference in relative fundamentality between two roles means that the more fundamental concept or the distinction between the two concepts is doing any `work' in any relevant sense.} It would then turn out, \emph{a posteriori}, that there is in practice, i.e.\ extensionally, an almost strict dichotomy between pure spacetime and pure matter. Keeping this conceptual distinction would then be useful as a basic categorisation principle. If there were many Jesusses and few or no mere humans and mere gods, it would not be as helpful to stick to those two categories as the most basic conceptual categories. Similarly, if the regions of pure spacetime theories (such as, perhaps, certain Palatini approaches to modified gravity \cite{olmo2011}) and pure matter theories (such as, perhaps, certain examples of WIMP dark matter) in the space of theories turn out to be small, the spacetime--matter distinction does not play as useful a role in the cartography of that space. 

All this relates to the issue of whether the spacetime--matter distinction earns its spurs as a theory-external, heuristic tool. Given the lessons learnt from SFDM, it seems that focusing on two separate spaces of theories has actually hindered theory development, as it has blinded us to potentially vast regions in the space of theories. The case may thus be worse than the spacetime--matter distinction having been neutral when it comes to its fruitfulness for theory development: letting go of the distinction may actively help us in exploring the space of theories. And even if there is a unique theory that describes our world and that theory contains only elements that are pure matter or pure spacetime, we can learn about that theory by studying how it differs from hybrid theories.

In a similar vein, a strict adherence to a DM/MG dichotomy has not been fruitful with regards to the collaboration between the DM community and the MG community. The second upshot of these companion papers is that it undermines the somewhat infamous hostility between these communities. Upon realisation that a field being dark matter does not exclude it being also an aspect of spacetime, and vice versa, it becomes clear that allowing that the other community may be on the right track does not necessitate one's own community being misguided. The lack of collaboration between the communities of course also depends on many other historical, social and economic factors---e.g.\ differences in training, differences in funding---and (resulting) differences in preferred guiding principles and theoretical virtues (e.g.\ accounts of scientific explanation). Hence, it is implausible that removing the conceptual incompatibility will by itself build a full bridge between the communities, but hopefully it opens the door, if just a little bit. A stronger focus on hybrid theories may help develop a trading zone \cite[Ch.9]{galison1997} between the DM and MG communities \cite{martenseditorial}.

%\textcolor{red}{maybe stuff about trading zones here. Sure, sociological and other factors play a role, but still, a strict dichotomy hasn't helped. Focusing on the whole space of theories, and in particular on other hybrid theories ,might help improve communication between the communities. Form a bridge, a trading zone.}

%Firstly, improving the communication between these two camps of physicists in a debate that is somewhat infamous for its polemic nature and lack of effective communication, for a variety of physics-related, funding-realted, ideological, psychological, sociological and historical reasons.

%Could these hybrid theories form a semantic bridge between the DM and MG communities---do they reveal, rather than ``mere mortar'', a full-blown creole/interlanguage that might eventually even lead to an independent community/ research field, similarly to how biochemistry arose from biology and chemistry \cite{collinsevansgorman} \cite[p.33-34,43]{galison2010}? 

The third upshot of these companion papers---besides the move to a single space of theories and the effects on the sociology of the DM and MG communities---concerns their implications for one of the largest debates in the philosophy of physics, that between substantivalism and relationalism about spacetime. As this debate presupposes a conceptual distinction between spacetime and matter, it becomes moot for theories to which any of the breakdown interpretations applies. On the (quietist---albeit not the agnosticist---version of the) consubstantiality interpretation it is claimed that nothing further can be said beyond the claim that the field under consideration is both (dark) matter and (modified) spacetime and that these roles are on a par. This includes quietism on the issue of relative fundamentality that the substantivalism--relationalism debate is concerned with. Could the fundamentalist interpretations save the debate? In fact, if the debate is being glossed over as being about the relative fundamentality of spacetime and matter, it may even seem that the fundamentalist interpretations \emph{just are} different positions within that debate: DM-fundamentalism is relationalism, MG-fundamentalism is monistic substantivalism (also known as supersubstantivalism, dualism is (dualistic) substantivalism, and neutral monism is either so-called `emergent spacetime \& matter' or so-called `super-emergent spacetime' \cite{martens2019}. 

However, if a single object is both spacetime and matter, it cannot be that, qua substance(s), this object is relatively fundamental or emergent to itself, as relative fundamentality is an irreflexive relation. Moreover, the debate concerns the relata of spatiotemporal relations: are they spacetime points or matter? DM-fundamentalism, for instance, is perfectly compatible with the view that a field such as $\Phi$ is a matter field that is both `living on' the fundamental spacetime points of an underlying manifold and instantiating, in some emergent regime, the role of a modification of the structure of that manifold. This does not seem to resemble relationalism. Finally, an important difference in consequence that traditionally distinguishes substantivalism from relationalism is that the former position considers vacuum solutions to be physically possible whereas the latter typically (but not always) judges them to be impossible. However, it is not even clear whether one can ask whether vacuum solutions are possible in theories with hybrid objects, since it is not clear in the first place what vacuum solutions are supposed to be in such theories. If one obtains a vacuum solution by removing all matter and including all spacetime structure, then $\Phi$ would have to be both included and excluded to obtain a vacuum solution of SFDM. A detailed evaluation of (the feasibillity of) the substantivalism--relationalism debate in the context of hybrid theories such as SFDM goes beyond the scope of this paper. Suffice it to say that in the worst case scenario the debate becomes ill-defined and in the best case scenario it becomes even more complex than it had already become in the context of modern physics.

Finally one may ask why these companion papers focused on SFDM and not on, say, GR? Does GR not stand a better chance at being empirically adequate? Should we not focus all our effort on interpreting GR? Several responses are appropriate here. First of all, suggesting that there are no empirical results speaking against GR is to beg the question against pure MG approaches, which claim that the empirical discrepancies that GR-advocates try to account for with DM actually show the empirical inadequacy of GR. That being said, SFDM is not a pure MG approach and we expect it to coincide with GR at $T \gg T_c$ (see the end of \hyperref[FDMintro]{Section~\ref{FDMintro}}). Hence, analysing SFDM does not mean that one is not also analysing aspects of GR. More generally, one can often learn more about theories, even familiar ones such as GR, by asking interpretative questions about other, neighbouring theories and comparing and contrasting the answers with the answers known from the familiar theories; at the very least it teaches us what is special about our mainstream theories and what is not. Finally, having a single theory, such as SFDM, that doesn't fit a strict DM/MG dichotomy, suffices to claim that there is a single space of theories that is irreducible to a space of spacetime theories and a space of matter theories---whether that theory is ultimately the correct description of our world does not change that. We have simply chosen SFDM because $\Phi$, being massive, is more clearly and less controversially a hybrid object than the metric in GR is (see e.g.~\hyperref[jesus]{Section~\ref{jesus}} and \cite[\S 5.1]{martenslehmkuhl1}). SFDM worsens the cracks in the spacetime--matter dichotomy that have already arisen from GR \cite[\S 2.2 \& 5.2]{martenslehmkuhl1}. %On Read's functionalism (\cite{read2018} \cite[section 2.2]{martenslehmkuhl1}), the Einstein metric carries energy\footnote{In the sense of energy being ascribable to a finite volume of spacetime; see matter criterion E in the prequel paper \cite{martenslehmkuhl1}.} in some solutions of GR, but not in others. Similarly, the Einstein metric satisfies the geodesic criterion (\S \ref{spacetime-matter}) in (many) non-rotating solutions of GR, but not, it has been argued, in for instance the G\"{o}del solution \cite{asenjohojman2017a,asenjohojman2017b,menonlinnemannread2018} \cite[section 5.2]{martenslehmkuhl1}. 
SFDM further teaches us that the categorisation of objects can vary across theories in the space of theories---compare the Einstein metric in GR to the Einstein metric in SFDM---and even within single solutions: $\Phi$ satisfies the spacetime criteria in the superfluid regime only. %Given what we have learnt from SFDM, are we still so sure that we understand plain old GR as well as we thought we did?

None of these reasons for considering SFDM is to say that we could not have chosen other theories to focus on. One simply has to start somewhere. Next steps in the cartography research programme could be to ask similar interpretational questions in different contexts, such as spin-2 gravity, Newton-Cartan theory, f(R) gravity and Jordan-Brans-Dicke theories, other hybrid DM/MG theories,\footnote{Read \emph{et al.}\ seem to suggest that TeVeS itself may be a hybrid theory \cite{readbrown2018}. For potential further, interesting case studies, see \cite{blanchet2008,zhao2008,bruneton2009,li2009,ho2010,ho2011,ho2012,cadoni2018,cadoni2019,scholz2020,ferreira2020,skordis2020}.} the cosmological constant, black holes, unified field theories, supersymmetric theories, and theories of quantum gravity\footnote{A particularly interesting example could be the dilaton in string theory. We would like to thank an anonymous reviewer for this suggestion.}. Within the cartography research programme, progress requires a continuous back and forth between analysis of individual theories and analysis of the space of theories. Another interesting follow-up would be to determine which modifications of SFDM would make $\Phi$ more or less of a dark matter field, and more or less of a modification of spacetime.\footnote{\label{multipleST}For instance, a very recent modification of SFDM by Berezhiani, Famaey \& Khoury \cite{berezhiani2017} couples photons not to $\tilde{g}_{\mu\nu}$ but to $g_{\mu\nu}$. \emph{Prima facie}, this undermines our claim that $\Phi$ is a modification of spacetime \cite[section 5.2]{martenslehmkuhl1}, and may seem to release the pressure on the matter--spacetime distinction by rendering $\Phi$ a pure matter field. On second thought, it may indicate either a) that there are several spacetimes---one experienced by photons, the other by the other luminous matter---or b) that both $g_{\mu\nu}$ and $\tilde{g}_{\mu\nu}$ are structural aspects of a single underlying spacetime; or c) that there is no spacetime at all in SFDM (which prompts the question of how to categorise $g_{\mu\nu}$ and $\tilde{g}_{\mu\nu}$); or d) that the strong geodesic criterion should be weakened to exclude light rays, i.e.\ it should be replaced by what we call the weak geodesic criterion \cite[section 3.2]{martenslehmkuhl1}. Neither option paints a simple picture of the spacetime--matter distinction. See also \cite{ferreira2019,ferreira2020}.} Extending the current context and analysing other contexts are however work for another day. We hope to have provided some useful tools---possible interpretations and arguments in favour of and against various versions of these interpretations---in exploring this map.

\section{Conclusion}

%\todo{Use the terminology in Dennis' suggested title: The breakdown of the distinction between dark matter and modifying gravity - and it’s resurrection?}

There is a single space of theories, which is not reducible to (just) a space of spacetime theories and another space of matter theories. This conclusion was reached via Superfluid Dark Matter Theory, which has helped highlight an important feature of the families of criteria for being matter and for being spacetime (which were reported in the prequel paper): they do not not form a strict dichotomy, but allow objects to satisfy criteria of either one family, of both, or of neither. The `cartography research programme' aims to analyse individual theories or pairs of theories in the space of theories in order to better understand the space of theories as a whole, which in turn helps us understand further individual theories and pairs of theories. This paper has provided a (non-exhaustive) chart of nine interpretations---structured into three groups---that may help navigate the space of theories. The chart was illustrated by applying it to the novel scalar field $\Phi$ within SFDM, which was concluded in the prequel paper to be as much (dark) matter as it could possibly be, and---at least in the superfluid regime---as much (of a modification of) spacetime as it could possibly be. Five of the nine interpretations have been ruled out for $\Phi$; at least one interpretation within each of the three groups remains on the table: the consubstantiality interpretation, DM-fundamentalism, (MG+DM)-fundamentalism (also known as dualism), and the weak breakdown interpretation. Although it is expected that finding a complete theory of SFDM may further narrow down the interpretational options, the arena where the ultimate fate of the DM/MG distinction and the broader matter--spacetime distinction, i.e.\ of our Democritean-Newtonian goggles, will be decided is likely the space of theories. We hope to have provided some useful cartographic tools---possible interpretations of Janus-faced theories and arguments in favour of and against these interpretations---towards this endeavour. Everyone is cordially invited to contribute to the cartography research programme. Physicists from the DM and MG communities are encouraged to consider focusing on a single space of theories as an overarching framework, and to use hybrid theories as a bridge across the gap between the communities. Philosophers of physics are encouraged to consider the interpretational questions voiced in these companion papers in different contexts, i.e.~for different points and pairs of (neighbouring) points in the space of theories, to consider the feasibility of the substantivalism--relationalism debate when it comes to hybrid theories, and to enrich the interpretational chart; in other words, to colour in the rough sketch of a map provided here.

\section*{Acknowledgements}

We would like to acknowledge support from the DFG Research Unit ``The Epistemology of the Large Hadron Collider'' (grant FOR 2063). Within this research unit we are particularly indebted to the other members of our `LHC, Dark Matter \& Modified Gravity' project team---Miguel \'{A}ngel Carretero Sahuquillo, Michael Kr\"{a}mer and Erhard Scholz---for invaluable and extensive discussions and comments on many iterations of this paper. We would furthermore like to thank Florian Boge, Jamee Elder, Alex Franklin, Tushar Menon, James Read, Katie Robertson, Kian Salimkhani, Michael St\"{o}ltzner and Adrian W\"{u}thrich for valuable discussions and comments, as well as the audiences of the Dark Matter \& Modified Gravity Conference (Aachen, Germany, 2019), the German Society for the Philosophy of Science Conference (Cologne, Germany, 2019) and the First Oxford-Notre Dame-Bonn Workshop on the Foundations of Spacetime Theories (Oxford, UK, 2019).

\bibliographystyle{unsrt}
\bibliography{bibdmmg}

\end{document}